\documentclass[
  journal=pasa,
  manuscript=research-article,
  year=202X,
  volume=XX,
]{cup-journal}

\usepackage{amsmath}
\usepackage[nopatch]{microtype}
\usepackage{booktabs}

\usepackage{xspace}
\usepackage[nolist]{acronym}
\usepackage{multicol}
\usepackage{supertabular}
\usepackage{multirow}
\usepackage{comment}

\usepackage{microtype,siunitx,booktabs}
\usepackage{makecell}
\usepackage{hyperref} 
\hypersetup{colorlinks,citecolor=blue,linkcolor=blue,urlcolor=blue}
\usepackage[nolist]{acronym}
\usepackage[version=3]{mhchem}
\usepackage{mathtools}
\usepackage{adjustbox}
\usepackage[normalem]{ulem}
\usepackage{xcolor}
\usepackage{comment}
\usepackage{amssymb}
\usepackage{upgreek}
\usepackage{amsfonts}
\usepackage{pifont} 
\usepackage{tikz} 
\usepackage{graphicx}
\usepackage{float}
\usepackage{footmisc}

\begin{acronym}[AWGN]
\acro{2MASS}{Two Micron All Sky Survey}
\acro{2MASX}{Two Micron All Sky Survey Extended Source Catalogue}
\acro{30Dor}[30 Dor]{30~Doradus}
\acro{AeReS}{Aegean Residual}
\acro{AGN}{active galactic nucleus}
\acrodefplural{AGN}{active galactic nuclei}
\acro{ATCA}{Australia Telescope Compact Array}
\acro{ATESP}{Australia Telescope ESO Slice Project}
\acro{ATNF}{Australia Telescope National Facility}
\acro{ATOA}{Australia Telescope Online Archive}
\acro{AT20G}{Australia Telescope 20 GHz Survey}
\acro{ASKAP}{Australian Square Kilometre Array Pathfinder}
\acro{ARC}{Australian Research Council}
\acro{BETA}{Boolardy Engineering Test Array}
\acro{BCG}{Brightest Cluster Galaxy}
\acro{BGG}{Brightest Group Galaxy}
\acro{BL}[BL Lac]{BL Lacertae Objects}
\acro{CASS}{CSIRO Astronomy and Space Science}
\acro{CABB}{Compact Array Broadband Back-end}
\acro{CHII}[{\sc CHii}]{Compact \textsc{Hii}}
\acro{CMB}{Cosmic Microwave Background}
\acro{CSIRO}{Australian Commonwealth Scientific and Industrial Research Organisation}
\acro{CSM}{circumstellar material}
\acro{CSS}{Compact Steep Spectrum}
\acro{DD}{double degenerate}
\acro{DECaLS}{Dark Energy Camera Legacy Survey}
\acro{DESI}{Dark Energy Spectroscopic Instrument}
\acro{DR3}{Data Release 3}
\acro{DS9}[\textsc{DS9}]{\textsc{SAOImage DS9}}
\acro{DSA}{diffusive shock acceleration}
\acro{DSS}{Digital Sky Survey}
\acro{EBHIS}{Effelsberg--Bonn H\,{\sc i} Survey}
\acro{EM}{Electromagnetic}
\acro{EMU}{Evolutionary Map of the Universe}
\acro{eROSITA}{extended R\"Ontgen Survey with an Imaging Telescope Array}
\acro{ev}[eV]{electronvolt\acroextra{: 1 eV $\approx 1.6 \times 10^{-19}$ J}}
\acro{FITS}[\textsc{Fits}]{Flexible Image Transport System}
\acro{FRB}{Fast Radio Bursts}
\acro{FSRQ}{Flat Spectrum Radio Quasars}
\acro{FWHM}{Full Width at Half-Maximum}
\acro{GASS}{Parkes Galactic All-Sky Survey}
\acro{GB}{giant branch}
\acro{GLEAM}{GaLactic and Extragalactic All-sky MWA Survey}
\acro{GRB}{Gamma-ray burst}
\acro{GPS}{Gigahertz Peak Spectrum}
\acro{HEMT}{High Electron Mobility Transistor}
\acro{H.E.S.S.}{High Energy Stereoscopic System}
\acro{HFP}[HFP]{High-Frequency Peaker}
\acro{HPBW}{Half Power Beam Width} 
\acro{HzRGs}{High Redshift Radio Galaxies}
\acro{pHFP}[pHFP]{Potential High Frequency Peaker}
\acro{HI}[H{\sc i}]{Neutral Atomic Hydrogen} 
\acro{HST}{\textit{Hubble Space Telescope}}
\acro{ICM}{intracluster medium}
\acro{IAU}{International Astronomical Union}
\acro{IFRSs}{Infrared Faint Radio Sources}
\acro{ISM}{Interstellar Medium}
\acro{IFRS}{Infrared Faint Radio Source}
\acro{IR}{infrared}
\acro{JY}[Jy]{Jansky\acroextra{, 1 Jy = $10^{-26} \times \mathrm{W~ m}^{-2}~\mathrm{Hz}^{-1}$}} 
\acro{LLS}{Largest Linear Size}
\acro{LFAA}{Low-Frequency Aperture Array}
\acro{LMC}{Large Magellanic Cloud}
\acro{LOFAR}{Low-Frequency Array}
\acro{LSO}[LSOs]{Large Scale Objects}
\acro{LBV}{Luminous Blue Variable}
\acro{MACHO}{Massive Astrophysical Compact Halo Objects}
\acro{MC}[MC]{Magellanic Cloud}    
\acro{MCELS}{Magellanic Cloud Emission Line Survey}
\acro{MS}{Magellanic Stream}
\acro{MW}{Milky Way}
\acro{MIRIAD}[\textsc{Miriad}]{Multichannel Image Reconstruction, Image Analysis and Display}
\acro{MIT}{Massachusetts Institute of Technology}
\acro{MOST}{Molonglo Observatory Synthesis Telescope}
\acro{MGPS-2}{second epoch Molonglo Galactic Plane Survey}
\acro{MQS}{Magellanic Quasars Survey}
\acro{MRC}{Molonglo Reference Catalogue of Radio Sources}
\acro{MWA}{Murchison Widefield Array}
\acro{NED}{NASA/IPAC Extragalactic Database}
\acro{NRAO}{National Radio Astronomy Observatory}
\acro{NSR}{Nova Super Remnant}
\acro{NVSS}{NRAO VLA Sky Survey}
\acro{OPAL}{Online Proposal Applications \& Links}
\acrodefplural{ORC}{Odd Radio Circles}
\acro{ORC}{Odd Radio Circle}
\acro{OVV}{Optically Violent Variable Quasars}    
\acro{PACS}{Photodetector Array Camera and Spectrometer}
\acro{PAF}{Phased Array Feed}
\acro{pc}{parsec\acroextra{: 1 pc $\simeq 3.09 \times 10^{16}$ m}}
\acro{PMN}{Parkes-MIT-NRAO}
\acro{PN}{Planetary Nebula}
\acro{PNe}{Planetary Nebulae}
\acro{PWN}{pulsar-wind nebula}
\acrodefplural{PWN}[PWNe]{pulsar-wind nebulae}
\acro{QSO}{Quasi-Stellar Object}
\acro{RA}{Right Ascension}
\acro{RFI}{Radio-Frequency Interference}
\acro{RMS}[RMS]{Root Mean Squared}
\acro{RM}{rotation measure}
\acro{SARAO}{South African Radio Astronomy Observatory}
\acro{SDSS}{Sloan Digital Sky Survey}
\acro{SED}{spectral energy distribution}
\acro{SGPS}{Southern Galactic Plane Survey}
\acro{SHS}{SuperCOSMOS H$\alpha$ Survey}
\acro{SI}[$\alpha$]{Spectral Index\acroextra{, $S \propto \nu^\alpha$}}
\acro{SKA}{Square Kilometre Array}
\acro{SMB}{Super Massive Blackholes}
\acro{SMC}{Small Magellanic Cloud}
\acro{SN}{supernova}
\acro{SUMSS}{Sydney University Molonglo Sky Survey}
\acro{TOPCAT}[\textsc{Topcat}]{Tool for OPerations on Catalogues And Tables}
\acro{USS}{Ultra Steep Spectrum}
\acro{YSOs}{young stellar objects}
\acro{WALLABY}{Widefield ASKAP L-band Legacy All-sky Blind surveY}
\acro{WBAC}{Wide-Band Analogue Correlator}
\acro{WD}{white dwarf}
\acro{WIFES}[WiFeS]{Wide-Field Spectrograph}
\acro{WISE}{Wide-Field Infrared Survey Explorer}
\acro{WR}{Wolf-Rayet}
\acro{VLA}{Very Large Array} 
\acro{VLBI}{Very Long Baseline Interferometry} 
\acro{VLSR}[\textbf{$v_{lsr}$}]{Velocity in the Line of Sight}
\acro{SMASH}{Survey of the Magellanic Stellar History}
\acro{SNR}{supernova remnant}
\acrodefplural{SNR}{supernova remnants}
\acro{SD}{single-degenerate}
\acro{SUMSS}{Sydney University Molonglo Sky Survey}
\acro{SMBH}{Super Massive Black Hole}
\acro{POSSUM}{Polarisation Sky Survey of the Universe's Magnetism}
\acro{CASDA}{CSIRO ASKAP Science Data Archive}
\acro{MIPS}{Multiband Imaging Photometer for Spitzer}
\acro{SN}{Supernova}
\acro{CC}{Core Collapse}
\acro{Cas A}{Cassiopeia A}
\acro{CARTA}{Cube Analysis and Rendering Tool for Astronomy}
\acro{IRAC}{Infrared Array Camera}
\acro{MIR}{Mid-Infrared}
\acro{SMGPS}{SARAO MeerKAT Galactic Plane Survey}
\end{acronym}

\newcommand{\ergcm}[1]{erg\,cm$^{-2}$\,s$^{-1}$}
\def\HI{\hbox{H{\sc i}}}

\newcommand{\D}{$^\circ$}

\newcommand{\farcm}{\mbox{\ensuremath{.\mkern-4mu^\prime}}}
\newcommand{\farcs}{\mbox{\ensuremath{.\!\!^{\prime\prime}}}}

\def\HI{\hbox{H{\sc i}}}

\def\arcmin{\hbox{$^\prime$}}
\def\arcsec{\hbox{$^{\prime\prime}$}}

\newcommand{\ujybm}{\,$\mu$Jy\,beam$^{-1}$}

\newcommand{\WISE}{{\it WISE}}
\newcommand{\Spitzer}{{\it Spitzer}}

\title{Stingrays in the radio sky: Two unusual diffuse radio relic sources in the direction of the Magellanic Stream}

\author{Z. J. Smeaton}
\affiliation{Western Sydney University, Locked Bag 1797, Penrith South DC, NSW 2751, Australia}
\email[Z. J. Smeaton]{19594271@student.westernsydney.edu.au

$\dag$Deceased on 3$^{\rm rd}$ July 2024}

\author{M. D. Filipovi\'c}
\affiliation{Western Sydney University, Locked Bag 1797, Penrith South DC, NSW 2751, Australia}

\author{B. S. Koribalski}
\affiliation{Australia Telescope National Facility, CSIRO, Space and Astronomy, PO Box 76, Epping, NSW 1710, Australia}
\alsoaffiliation{Western Sydney University, Locked Bag 1797, Penrith South DC, NSW 2751, Australia}

\author{M. Sasaki}
\affiliation{Dr Karl Remeis Observatory, Erlangen Centre for Astroparticle Physics, Friedrich-Alexander-Universit\"{a}t Erlangen-N\"{u}rnberg, Sternwartstra{\ss}e 7, 96049 Bamberg, Germany}

\author{R. Z. E. Alsaberi}
\affiliation{Faculty of Engineering, Gifu University, 1-1 Yanagido, Gifu 501-1193, Japan}
\alsoaffiliation{Western Sydney University, Locked Bag 1797, Penrith South DC, NSW 2751, Australia}

\author{A. C. Bradley}
\affiliation{Western Sydney University, Locked Bag 1797, Penrith South DC, NSW 2751, Australia}

\author{E. J. Crawford}
\affiliation{Western Sydney University, Locked Bag 1797, Penrith South DC, NSW 2751, Australia}

\author{S. Dai}
\affiliation{Western Sydney University, Locked Bag 1797, Penrith South DC, NSW 2751, Australia}
\alsoaffiliation{Australia Telescope National Facility, CSIRO, Space and Astronomy, PO Box 76, Epping, NSW 1710, Australia}

\author{N.~Gupta}
\affiliation{Australia Telescope National Facility, CSIRO, Space and Astronomy, PO Box 1130, Bentley, WA 6102, Australia}

\author{F. Haberl}
\affiliation{Max-Planck-Institut f\"ur extraterrestrische Physik, Gie\ss enbachstra\ss e 1, 85748 Garching, Germany}

\author{A. M. Hopkins}
\affiliation{School of Mathematical and Physical Sciences, 12 Wally's Walk, Macquarie University, NSW 2109, Australia}

\author{T. H. Jarrett$^\dag$}
\affiliation{Department of Astronomy, University of Cape Town, Private Bag X3, Rondebosch 7701, South Africa}
\alsoaffiliation{Western Sydney University, Locked Bag 1797, Penrith South DC, NSW 2751, Australia}

\author{S. Lazarevi\'c}
\affiliation{Western Sydney University, Locked Bag 1797, Penrith South DC, NSW 2751, Australia}
\alsoaffiliation{Australia Telescope National Facility, CSIRO, Space and Astronomy, PO Box 76, Epping, NSW 1710, Australia}
\alsoaffiliation{Astronomical Observatory, Volgina 7, 11060 Belgrade, Serbia}

\author{D. Leahy}
\affiliation{Department of Physics and Astronomy, University of Calgary, Calgary, Alberta, T2N IN4, Canada}

\author{P. Macgregor}
\affiliation{Western Sydney University, Locked Bag 1797, Penrith South DC, NSW 2751, Australia}
\alsoaffiliation{Australia Telescope National Facility, CSIRO, Space and Astronomy, PO Box 76, Epping, NSW 1710, Australia}

\author{G. Rowell}
\affiliation{School of Physics, Chemistry and Earth Sciences, The University of Adelaide, Adelaide 5005, Australia}

\author{S. S. Shabala}
\affiliation{School of Natural Sciences, Private Bag 37, University of Tasmania, Hobart, TAS 7001, Australia}

\author{D. Uro\v{s}evi\'c}
\affiliation{Department of Astronomy, Faculty of Mathematics, University of Belgrade, Studentski trg 16, 11000 Belgrade, Serbia}

\author{J. Th. van Loon}
\affiliation{Lennard-Jones Laboratories, Keele University, ST5 5BG, UK}

\author{T. Vernstrom}
\affiliation{Australia Telescope National Facility, CSIRO, Space and Astronomy, PO Box 1130, Bentley, WA 6102, Australia}
\alsoaffiliation{International Centre for Radio Astronomy Research (ICRAR), The University of Western Australia, 35 Stirling Highway}


\begin{document}

\begin{abstract}

We present the discovery of two extended, low surface brightness radio continuum sources, each consisting of a near-circular body and an extended tail of emission, nicknamed Stingray~1 (ASKAP~J0129--5350) and Stingray~2 (ASKAP~J0245--5642). Both are found in the direction of the \ac{MS} and were discovered in the \ac{ASKAP} \ac{EMU} survey at 944\,MHz. We combine the \ac{ASKAP} data with low-frequency radio observations from the \ac{GLEAM} to conduct a radio continuum analysis. 
We explore both Galactic/near Galactic scenarios, including runaway or circumgalactic \acp{SNR} and parentless \acp{PWN}, and extragalactic scenarios including radio \acp{AGN}, dying radio galaxies, galaxy clusters, galaxy pairs or groups, head-tail radio galaxies, and \acp{ORC}, as well as the possibility that the morphology is due to a chance alignment. The Stingrays exhibit non-thermal emission with spectral indices of $\alpha=-0.89\pm0.09$ for Stingray~1 and $\alpha=-1.77\pm0.06$ for Stingray~2. We find that none of the proposed scenarios can explain all of the observed properties, however we determine it most likely that their shape is caused by some kind of complex environmental interaction. The most likely scenario from the available data is that of a head-tail radio galaxy, but more data is required for a definitive classification.

\end{abstract}

\acresetall

\section{Introduction}
 \label{sec:1}

The \ac{EMU} \citep[][AS201]{Norris2011, Norris2021, 2025PASA...42...71H} project is a large-scale radio survey conducted with the \ac{ASKAP}. \ac{ASKAP}'s modern phased array feed technology allows it to achieve an instantaneous field of view of $\sim$30~deg$^2$ enabling wide-field imaging surveys \citep{2008ExA....22..151J,2021PASA...38....9H}. By using the complete array of 36 \ac{ASKAP} antennas, \ac{EMU} is currently imaging the entire southern sky with better sensitivity ($\sim$30~$\mu$Jy\,beam$^{-1}$) and resolution (15\arcsec) than previous all-sky surveys. 

The \ac{EMU} survey is currently ongoing and its improved sensitivity has helped to reveal new low radio surface-brightness objects. A recent study by~\citet[]{Ball2023} has shown \ac{ASKAP}'s ability to uncover a previously undiscovered population of low surface-brightness Galactic \acp{SNR}, and this is also demonstrated by several other \ac{SNR} and \ac{SNR} candidate studies and discoveries (e.g., G278.92+1.35; Diprotodon \citep{2024PASA...41..112F}, G181.1--9.5~\citep{2017A&A...597A.116K}, G288.8--6.3; Ancora \citep{2023AJ....166..149F,2024A&A...684A.150B}, G32.9--0.5; Perun \citep{Smeaton2024a}, G308.73+1.38; Raspberry \citep{2024RNAAS...8..107L}, G312.65+2.87; Unicycle \citep{2024RNAAS...8..158S}; Teleios \citep{2025PASA...42..104F}), including the first circumgalactic \ac{SNR} J0624--6948 \citep{2022MNRAS.512..265F, 2025Sasaki}. Additionally, \ac{ASKAP} was used to discover a new class of low surface-brightness radio sources known as \acp{ORC} whose nature and origin is still under investigation \citep{2021PASA...38....3N, 10.1093/mnrasl/slab041, 2022MNRAS.513.1300N}. Also, using \ac{ASKAP}, we found a very unusual \ac{AGN} with recollimated jets in nearby NGC\,2663 \citep{2022MNRAS.516.1865V} as well as peculiar galaxy pair PKS~2130-538 within Abell~3785 cluster \citep{Velovic2023}. \ac{ASKAP}'s sensitivity was also able to uncover some unusual faint, filamentary structures in the Abell S1136 galaxy cluster \citep{2024PASA...41...50M}. \\

\begin{figure*}
    \centering
    \includegraphics[width=14.2cm]{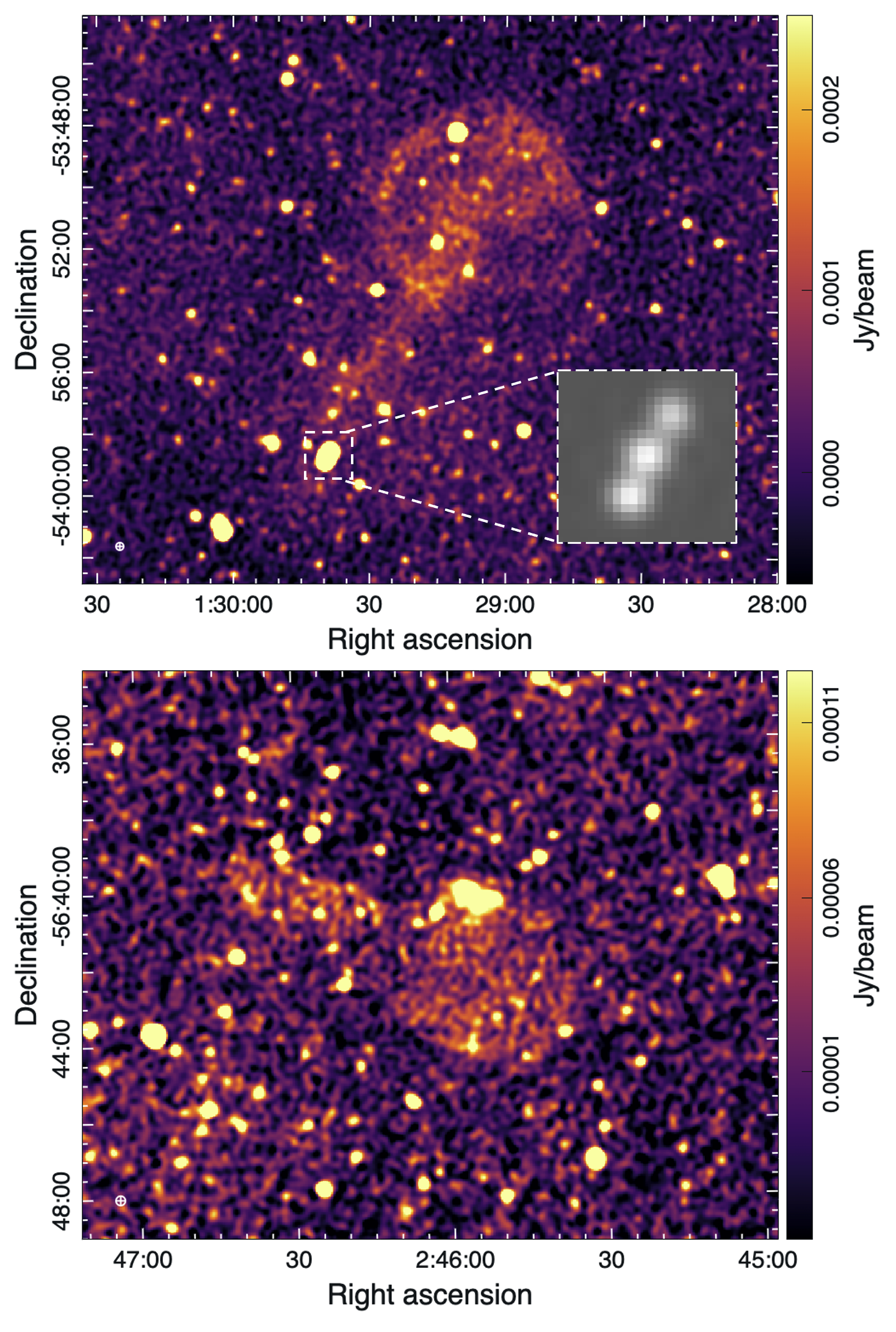}
    \caption{ASKAP EMU 944\,MHz radio continuum images of two peculiar sources, named Stingray~1 (top) and Stingray~2 (bottom). Both images use linear scaling and have a restoring beam of 15\arcsec\ which is shown in the bottom left corners. The \ac{RMS} noise sensitivities are $\sigma$ = 25\ujybm\ for Stingray~1 and $\sigma$ = 30\ujybm\ for Stingray~2. The inset image in Stingray~1 shows the high-resolution (7\farcs7 $\times$ 7\farcs0) \ac{ASKAP} image of the double-lobed radio galaxy associated with the galaxy WISEA~J012939.26--535841.0 (see Sect.~\ref{subsubsec:AGN}).}
    \label{fig:ASKAP}
\end{figure*}

In this paper, we add to this list of interesting \ac{ASKAP} discoveries two low surface-brightness diffuse radio sources. We nickname these objects ``Stingrays'' due to their unusual head-tail radio shape (see Figure~\ref{fig:ASKAP}) resembling the animal. Another interesting feature is the Stingray's location in the direction of the \ac{MS} which consists of \HI\ gas that trails behind the \acp{MC}, a pair of nearby interacting dwarf galaxies. The origin of the \ac{MS} is a topic of active investigation~\citep{Lucchini_2021, 10.1093/mnras/stac1640, Chandra_2023, Lucchini_2024}. 

We conduct a radio continuum analysis using the available data to attempt to identify the nature of the Stingrays. We investigate multiple possibilities, such as runaway \acp{SNR} within the \ac{MS}, circumgalactic \acp{SNR}, \acp{PWN}, as well as further extragalactic scenarios such as radio \acp{AGN}, dying radio galaxies, galaxy clusters, galaxy pairs or groups, head-tail radio galaxies, and \acp{ORC}, as well as the possibility of a chance alignment of the different morphological components.

\begin{table*}
\centering
\caption{Details of the main radio continuum observations used in this work. --- Notes: The \ac{ASKAP} observation dates are different for each object -- we use (S1) for Stingray~1} and (S2) for Stingray~2 -- and the \ac{MWA} beam size and \ac{RMS} are image-dependent. The HI4PI data is merged data from GASS and EBHIS. The \ac{RMS} units are mJy\,beam$^{-1}$ for all data except for HI4PI, which is mK.
\vskip.25cm
\begin{tabular}{|ccccccc|} 
\hline
    Telescope & Project & Observing Date & Frequency & Beam Size & Pixel Size & RMS \\
    & & & (MHz) & (arcsec) & (arcsec) & \\
\hline
    & & & 88 & 303$\times$296 (S1) &  & $\sim$50 \\
    & &  &  & 308$\times$297 (S2) &  & \\
    & &  & 118 & 222$\times$216 (S1) &  & $\sim$30 \\
    & &  &     & 222$\times$216 (S2) &  & $\sim$25 \\
    MWA & GLEAM & 9 Aug 2013--18 Jun 2014 & 155 & 173$\times$167 (S1) & 24 & $\sim$15 \\
    & &  &  & 170$\times$163 (S2) &  & $\sim$12 \\
    & &  & 200 & 145$\times$138 (S1) &  & $\sim$5\\ 
    & &  &  & 143$\times$134 (S2) &  & $\sim$3 \\ 
\hline
    & EMU & 17 May 2023 and 22 Mar 2024 (S1) & 944 & 15 & 2 & 0.025 \\
    ASKAP & EMU & 17 May 2023 (S1) & 944 & $7.7 \times 7.0$ & 2 & 0.055 \\ 
    & EMU & 13 May 2023 (S2) & 944 & 15 & 2 & 0.030 \\
    & WALLABY & 21 Oct 2024 and 29 Oct 2024 (S1) & 1368 & $\sim$8 & 2 & 0.035 \\
    & WALLABY & 21 Oct 2024 and 29 Oct 2024 (S1) & 1368 & 15 &  2 & 0.025--0.030 \\
\hline
    Parkes & HI4PI & Jan 2005--Nov 2006 (GASS) & 1420 & 974 & 300 & 43 \\ 
\hline
\end{tabular}
\label{tab:observations}
\end{table*}

The paper is structured as follows; our observations and data are discussed in Sect.~\ref{sec:obs and data reduction}, analysis methods and results are in Sect.~\ref{sec:results}, a broad theoretical discussion is in Sect.~\ref{sec:discussion}, and our conclusions are in Sect.~\ref{sec:conclusion}.


\section{Observations and Data Reduction}
 \label{sec:obs and data reduction}

This study primarily uses radio continuum data from the \ac{ASKAP} \ac{EMU} sky survey and the \ac{GLEAM} as well as \HI\ data from the HI4PI survey. See Table~\ref{tab:observations} for a summary of the observational properties. We searched for corresponding diffuse emission at multiple other frequencies, outlined in Section~\ref{sec:results}. The primary image analysis software used was the \ac{CARTA} \citep{carta}.

\subsection{ASKAP}
\label{subsec:ASKAP}

\subsubsection{EMU}
\label{subsubsec:EMU}

The areas of sky where the two Stingrays are located were observed as part of the large-scale \ac{EMU} project \citep{2025PASA...42...71H, Norris2011, Norris2021} (AS201) using the complete set of 36 \ac{ASKAP} antennas at a central frequency of 943.5\,MHz and a bandwidth of 288\,MHz. Stingray~1 (ASKAP~J0129--5350, Figure~\ref{fig:ASKAP} top) was observed during two different scheduling blocks, SB50048 and SB60320, while Stingray~2 (ASKAP~J0245--5642, Figure~\ref{fig:ASKAP} bottom) was observed in SB49990. The total integration times are 20~h for Stingray~1 and 10~h for Stingray~2. The data were processed using the standard ASKAPsoft pipeline, including multi-scale cleaning, self-calibration, and multi-frequency synthesis imaging \citep{askapsoft_2019ascl.soft12003G}. The images were downloaded from the CSIRO \ac{ASKAP} Science Data Archive (CASDA)\footnote{\url{https://research.csiro.au/casda}}.

The \ac{ASKAP} \ac{EMU} data consists of Stokes~$I$ and $V$ images. There is no detection in Stokes~V, and so only the Stokes~I images are used in our analysis. We use both the convolved and high resolution Stokes~I images which are produced by the standard ASKAPsoft pipeline \citep{askapsoft_2019ascl.soft12003G}. Both these images are generated from the same calibrated dataset, but have different imaging parameters applied. The high resolution images are made using uniform weighting, resulting in beam sizes of typically $\sim$7\arcsec$-$9\arcsec\ and increased noise. The convolved images are made using robust = 0 weighting and are restored using a 15\arcsec\ beam.
Further technical details can be found in \citet{2025PASA...42...71H}. The convolved images are the recommended science data products as they allow for accurate flux density measurements over the entire \ac{EMU} survey.

We therefore use the convolved radio continuum images, shown in  Figure~\ref{fig:ASKAP}, for any quantitative measurements such as flux density estimates, and hereafter the \ac{EMU} image will refer to the 944\,MHz convolved image unless otherwise stated. Since Stingray~1 is found in two different observations, these images were combined with the \textsc{imcomb} function in the \ac{MIRIAD} software package \citep{Sault1995} using equal weighting. The resulting image is shown in  Figure~\ref{fig:ASKAP} (top); the \ac{RMS} noise near Stingray~1 is 25\ujybm. The inset shows the high-resolution radio continuum image of the source WISEA J012939.26--535841.0 (see Section~\ref{subsubsec:AGN}) with a synthesized beam of 7\farcs7$\times$7\farcs0 and an \ac{RMS} of $\sigma$ = 55\ujybm. The bottom panel shows Stingray~2 in the convolved Stokes~$I$ image with a resolution of 15\arcsec\ and an \ac{RMS} of $\sigma$ = 30\ujybm.


\subsubsection{WALLABY}
\label{subsubsec:WALLABY}

Stingray~1 is also detected in 1.4~GHz radio continuum images from the \ac{WALLABY} (AS202)~\citep{2020Ap&SS.365..118K, 2022PASA...39...58W}, while the field containing Stingray~2 has not yet been observed. The central observing frequency for WALLABY is 1367.5\,MHz; due to RFI the bandwidth is limited to 144\,MHz. Stingray~1 is detected in two separate scheduling blocks, SB67270 and SB66917, with a combined integration time of 16~h. The data is processed similar to the \ac{EMU} data using the ASKAPsoft pipeline \citep{askapsoft_2019ascl.soft12003G} and the radio continuum images were downloaded from CASDA. Both images have angular  resolutions of $\sim$8\arcsec\ (8.3\arcsec$\times$7.7\arcsec, P.A.\,=\,74.9\D\ for SB67270, and 8.2\arcsec$\times$7.8\arcsec, P.A\,=\,81.4\D\ for SB66917) and were combined following the same process as for the \ac{EMU} images. We measure an \ac{RMS} noise sensitivity of $\sim$30--35\,$\mu$Jy\,beam$^{-1}$ near Stingray~1. To increase sensitivity to diffuse emission, we convolve the \ac{WALLABY} image to 15\arcsec\ resolution, matching that of the convolved \ac{EMU} image. This convolved \ac{WALLABY} image has an \ac{RMS} noise sensitivity of $\sim$25$-$30\,$\mu$Jy\,beam$^{-1}$ and is shown in the Appendix.

\subsection{MWA}
\label{MWA}
The sky region containing Stingrays~1 \& 2 was also observed with the Murchison Widefield Array (MWA) telescope as part of the \ac{GLEAM} survey \citep{2015PASA...32...25W, 2017MNRAS.464.1146H, 2019PASA...36...47H, 2019PASA...36...48H}. The data consists of 4 observations at the central frequencies of 88, 118, 155, and 200\,MHz. The image bandwidth is 30\,MHz, apart from the 200\,MHz observations which have a bandwidth of 60\,MHz. The size of the restoring beams and measured \ac{RMS} noise levels are listed in Table~\ref{tab:observations}.

\subsection{HI4PI}
\label{HI4PI}

We use \HI\ spectral line data from the HI4PI survey, which consists of data from the \ac{EBHIS} and the \ac{GASS}, as outlined in \cite{HI4PI}. Due to their location, the Stingrays are not observable by the Effelsberg telescope, but are covered by the Parkes telescope. The angular resolution is 16\farcm2 and the average \ac{RMS} noise for the final data is 43~mK per channel with a channel width of 1.29~km\,s$^{-1}$.

\section{Results}
\label{sec:results}

The Stingrays were discovered in the \ac{ASKAP} \ac{EMU} main survey when examining the 15\arcsec-resolution radio continuum images by eye while searching for interesting emission features. Following their discovery, we inspected images at multiple other frequencies, including infrared (\WISE\ and \Spitzer), optical (\ac{SDSS}, \ac{SHS}, and \ac{DECaLS}), X-ray (eROSITA), as well as searching in $\gamma$-ray catalogues \citep[\textit{Fermi} 4FGL-DR4,][]{2022ApJS..260...53A}. There are infrared, optical, and X-ray point sources that appear within the areas of the two Stingrays, however there is no diffuse emission at any other frequency that appears to correlate with the radio emission. Some of these point sources are investigated as potential host galaxies in Section~\ref{subsec:extragalactic}.

\subsection{Radio continuum}
\label{subsec:radio continuum}

\subsubsection{Morphology}
\label{subsubsec:morphology}

\begin{table*}
    \centering
    \caption{General morphological characteristics of the Stingrays including position, angular size (diameter for the circle region), and position angle (measured clockwise from North). The circle, tail, and total sections are defined by the regions outlined in Section~\ref{subsubsec:morphology}.}
    \vskip.25cm
    \begin{tabular}{|lc|ccc|ccc|} \hline
        Properties & Units & \multicolumn{3}{c|}{Stingray~1} & \multicolumn{3}{c|}{Stingray~2} \\ \cline{3-8}
         &  & Circle & Tail & Total Length & Circle & Tail & Total Length \\ \hline
        RA(J2000) & h:m:s & 01:29:07.0 & 01:29:32.5 & -- & 02:45:54.7 & 02:46:26.9 & -- \\
        Dec(J2000) & d:m:s & --53:50:46.7 & --53:56:34.7 & -- & --56:42:06.0 & --56:40:15.2 & -- \\
        Angular Size & arcmin & 7.0 & 6.8$\times$1.6 & 13.8 & 4.8 & 4.8$\times$1.2 & 9.6 \\
        Position Angle & degr & -- & 150 & -- & -- & 76 & -- \\ \hline
    \end{tabular}
    \label{tab:summary}
\end{table*}

We find two extended, very low-surface brightness radio sources with nearly matching, peculiar morphologies in the \ac{EMU} data (see Figure~\ref{fig:ASKAP}). Each consists of a near-circular body and extended tail, spanning several arcminutes in size. Optical images of both sources show a number of possibly associated galaxies. Interestingly, both objects were found relatively close to each other and in the direction of the Magellanic Stream.

To determine the angular sizes of the Stingrays, we fit two regions around each Stingray using the astronomical imaging software \ac{CARTA}. These regions are set by eye around the emission visible in the \ac{EMU} images and consist of a circular and a tail region for each Stingray. The angular dimensions of these regions are given in Table~\ref{tab:summary}. The total length of each Stingray is measured from the tip of the tail to the far side of the circle section {\bf (the far side being the edge of the circle opposite to the tail region)}. Overall, both Stingrays display a similar morphology with a circular head and extended tail, and Stingray~1 is $\sim$1.4 times larger than Stingray~2 in terms of total length.

There are several radio point sources within both Stingrays. We perform source finding using the \textsc{Aegean}\footnote{\url{https://github.com/PaulHancock/Aegean}} and Background and Noise Estimator (\textsc{BANE}) software of~\cite{2012MNRAS.422.1812H,2018PASA...35...11H}, then cross-match them with the astronomical databases, SIMBAD\footnote{\url{https://simbad.cds.unistra.fr/simbad/}}~\citep{2000A&AS..143....9W} and \ac{NED}\footnote{The NASA/IPAC Extragalactic Database (NED) is funded by the National Aeronautics and Space Administration and operated by the California Institute of Technology.
~\url{https://ned.ipac.caltech.edu/}}. 
Most of the sources appear in these databases and some have redshifts. These sources, and the possibility of an association with the Stingrays, is discussed further in Section~\ref{subsec:extragalactic}. 


\subsubsection{Flux density}
\label{subsubsec:flux}

\begin{figure*}
    \centering
    \includegraphics[width=\textwidth]{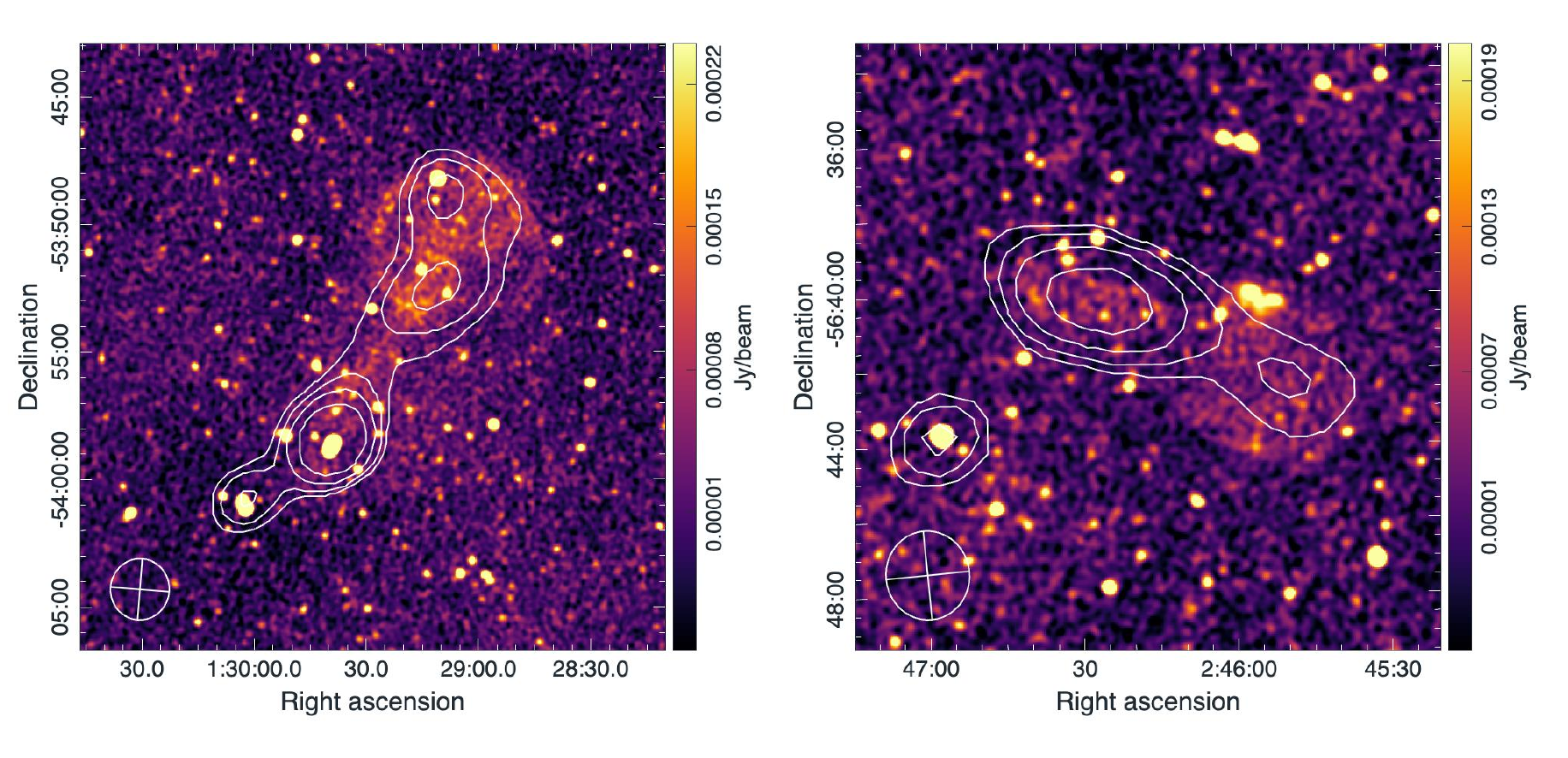}
    \caption{ASKAP EMU 944\, MHz radio continuum images of Stingray~1 (left) and Stingray~2 (right) overlaid with GLEAM 200\,MHz contours. Both images use linear scaling. The background image is the same \ac{EMU} image as shown in Figure~\ref{fig:ASKAP}. The contours are from the 200\,MHz \ac{MWA} image with a restoring beam of 145\arcsec$\times$138\arcsec\ and \ac{RMS} noise sensitivity of $\sigma$ = 4.5\,mJy beam$^{-1}$ for Stingray~1 (left), and a restoring beam of 143\arcsec$\times$134\arcsec\ and \ac{RMS} noise sensitivity of $\sigma$\,=\,3.0\,mJy beam$^{-1}$ for Stingray~2 (right) (see Table~\ref{tab:observations}). Contour levels are at $\sim$5, 7, 10, and 15 $\sigma$ for each image and the \ac{MWA} beam sizes are shown in the bottom left corners}.
    \label{fig:MWA}
\end{figure*}

We measure the flux density of the Stingrays with the imaging software \ac{CARTA} using the regions as defined in Section~\ref{subsubsec:morphology} and Table~\ref{tab:summary}. For the \ac{EMU} images, the convolved images are used due to their self-consistent flux density measurements (see Section~\ref{subsubsec:EMU}), and we remove the point sources using the \ac{AeReS} tool in the Aegean software package. \ac{AeReS} was unable to properly remove the two brightest point sources and so these were subtracted manually to obtain flux densities both with and without point sources (Table~\ref{tab:measured fluxes}). We refer to these flux densities throughout the text as the total flux density (total flux density from the regions including point sources), diffuse emission (total flux density from the regions once the point sources have been subtracted), and the point source flux density. For the \ac{WALLABY} data, the signal-to-noise is too low to be able to accurately measure any component other than the circular section's emission. We are unable to remove the point sources for the \ac{MWA} images due to the lower resolution (Figure~\ref{fig:MWA}). 

All images are then convolved to the largest beam (at 88~MHz) before flux extraction (the \ac{MWA} 88~MHz image has a  beam size of 303\arcsec$\times$296\arcsec\ for Stingray~1 and 308\arcsec$\times$297\arcsec\ for Stingray~2 and a pixel size of 24\arcsec\ for both; see Table~\ref{tab:observations}) to ensure consistent scaling as described in \citet[][Section 2.4]{2019PASA...36...48H}. Errors are estimated as $\sim$10\% following a similar process as~\citet[]{Filipovic2022}.

\begin{table*}
    \centering
    \caption{Measured flux densities of the Stingrays at all radio frequencies. All images were convolved to the lowest resolution for each object. For \ac{ASKAP}, three flux measurements are given as described in Section~\ref{subsubsec:flux}: total (including point sources), diffuse (excluding point sources), and point sources (the difference of total and diffuse). 
    Errors are taken as $\sim$10\% with a minimum error of $\pm$ 1~mJy.}
    \vskip.25cm
    \begin{tabular}{|lc|ccc|ccc|} \hline
     &  & \multicolumn{6}{c|}{$S_{\text{I}}$ (mJy)} \\ \cline{3-8}
     Telescope & $\nu$ & \multicolumn{6}{c|}{Stingray~1} \\ \cline{3-8}
     & (MHz) & \multicolumn{3}{c|}{Circle} & \multicolumn{3}{c|}{Tail}\\ \hline
      & 88 & \multicolumn{3}{c|}{439$\pm$44} & \multicolumn{3}{c|}{358$\pm$36}\\
     MWA & 118 & \multicolumn{3}{c|}{298$\pm$30} & \multicolumn{3}{c|}{253$\pm$25}\\
     & 155 & \multicolumn{3}{c|}{208$\pm$21} & \multicolumn{3}{c|}{180$\pm$18}\\
     & 200 & \multicolumn{3}{c|}{205$\pm$21} & \multicolumn{3}{c|}{143$\pm$14}\\ \hline
      &  & Total & Diffuse & Point & Total & Diffuse & Point\\
      \ac{ASKAP} &  &  &  & sources &  &  & sources\\ \cline{3-8}
      & 944 & 54$\pm$5 & 36$\pm$4 & 18$\pm$9 & 52$\pm$5 & 23$\pm$2 & 29$\pm$7\\ 
      &  1368 & 31$\pm$3 & 17$\pm$2 & 14$\pm$5 & -- & -- & -- \\ \hline \hline
      &  & \multicolumn{6}{c|}{Stingray~2} \\ \cline{3-8}
      &  & \multicolumn{3}{c|}{Circle} & \multicolumn{3}{c|}{Tail}\\ \hline
      & 88 & \multicolumn{3}{c|}{156$\pm$16} & \multicolumn{3}{c|}{273$\pm$27} \\
     MWA & 118 & \multicolumn{3}{c|}{90$\pm$9} & \multicolumn{3}{c|}{178$\pm$18}\\
     & 155 & \multicolumn{3}{c|}{52$\pm$5} & \multicolumn{3}{c|}{105$\pm$11} \\
     & 200 & \multicolumn{3}{c|}{39$\pm$4} & \multicolumn{3}{c|}{55$\pm$6} \\ \hline
      &  & Total & Diffuse & Point & Total & Diffuse & Point\\
      \ac{ASKAP} & 944 &  &  & sources &  &  & sources\\ \cline{3-8}
      &  & 7$\pm$1 & 4$\pm$1 & 3$\pm$2 & 5$\pm$1 & 2$\pm$1 & 3$\pm$2 \\ \hline
    \end{tabular}
    \label{tab:measured fluxes}
\end{table*}

\begin{table*}
    \centering
    \caption{Calculated scaled and diffuse flux densities of the Stingrays assuming a spectral index $\alpha$\,=\,--0.7 for extragalactic point sources. Scaled flux densities are the sum of the point source flux densities measured from the \ac{EMU} data (Table~\ref{tab:measured fluxes} `Point sources' column) scaled to \ac{MWA} frequencies using the assumed spectral index $\alpha$\,=\,$-$0.7. Diffuse flux densities are the difference between the total MWA flux densities from Table~\ref{tab:measured fluxes} and the scaled point source flux densities. Errors are taken as $\sim$10\%. The spectral index values calculated using these scaled fluxes and the fluxes from Table~\ref{tab:measured fluxes} are shown in the bottom two rows.}
    \vskip.25cm
    \begin{tabular}{|c|cc|cc|cc|cc|} \hline
         & \multicolumn{8}{c|}{$S_{\text{I}}$ (mJy)} \\ \cline{2-9}
        $\nu$ & \multicolumn{4}{c|}{Stingray~1} & \multicolumn{4}{c|}{Stingray~2} \\ \cline{2-9}
        (MHz) & \multicolumn{2}{c|}{Circle} & \multicolumn{2}{c|}{Tail} & \multicolumn{2}{c|}{Circle} & \multicolumn{2}{c|}{Tail} \\ \cline{2-9}
        & Scaled & Diffuse & Scaled & Diffuse & Scaled & Diffuse & Scaled & Diffuse \\ \hline
        88 & 95$\pm$10 & 344$\pm$34 & 153$\pm$15 & 205$\pm$21 & 16$\pm$2 & 140$\pm$14 & 16$\pm$2 & 257$\pm$26 \\
        118 & 77$\pm$8 & 221$\pm$22 & 124$\pm$12 & 129$\pm$13 & 13$\pm$1 & 77$\pm$8 & 13$\pm$1 & 165$\pm$17 \\
        155 & 64$\pm$6 & 144$\pm$14 & 103$\pm$10 & 77$\pm$8 & 11$\pm$1 & 41$\pm$4 & 11$\pm$1 & 94$\pm$9 \\
        200 & 53$\pm$5 & 152$\pm$15 & 86$\pm$9 & 57$\pm$6 & 9$\pm$1 & 30$\pm$3 & 9$\pm$1 & 46$\pm$5 \\ \hline
        $\alpha$ & \multicolumn{2}{c|}{-1.00$\pm$0.07} & \multicolumn{2}{c|}{-0.86$\pm$0.15} & \multicolumn{2}{c|}{-1.45$\pm$0.10} & \multicolumn{2}{c|}{-2.08$\pm$0.05} \\ \cline{2-9}
        & \multicolumn{4}{c|}{-0.89$\pm$0.09} & \multicolumn{4}{c|}{-1.77$\pm$0.06} \\ \hline
    \end{tabular}
    \label{tab:scaled fluxes}
\end{table*}

\subsubsection{Spectral index}
\label{subsubsec:spec index}

\acp{SNR} have theoretical and observed average spectral index values of $\alpha\ = -0.5$ due to synchrotron emission from the acceleration of charged particles in their expanding shockwave~\citep{Bell1978,2017ApJS..230....2B,2023MNRAS.518.2574B,2024MNRAS.529.2443C}. For the \ac{MC} \ac{SNR} population, observational studies show a mean value of $\alpha\approx-0.5$, with a distribution range $-0.9<\alpha<0$~\citep{2017ApJS..230....2B, 2019A&A...631A.127M}. Extragalactic radio sources typically have an average observed spectral index of $\alpha\sim-0.7$ to $\alpha\sim-0.8$, measured from independent large-scale radio studies~\citep{condon1984cosmological,2003MNRAS.342.1117M,2021MNRAS.506.3540P, 2021MNRAS.507.2885F}, however the range of possible spectral indices is far wider and can extend from ultra-steep spectrum sources~\citep{2014MNRAS.439..545C, 2018MNRAS.477..578C} to flat and inverted spectrum sources~\citep{2021MNRAS.507.2885F, 2022Ap&SS.367...61B, 2023MNRAS.519.4902S}. 


\paragraph{Total Spectral Index: }

We are able to separate the point source emission from the diffuse emission in the \ac{ASKAP} images, however we are unable to do so in the \ac{GLEAM} images due to the poorer resolution. As these are likely extragalactic radio sources with spectral indices $\sim$\,--0.7 to --0.8, they may contribute significantly at lower frequencies and result in a significant overestimate of the \ac{GLEAM} flux densities. 
Due to the 
similar observing frequencies between the \ac{EMU} and \ac{WALLABY} data, it is not possible to accurately measure the spectral indices for many of the point sources within Stingray~1. We are unable to measure any point source spectral indices in Stingray~2 due to the lack of \ac{WALLABY} data for this object. Therefore, we assume an average spectral index of $\alpha\,=\,-0.7${\citep{condon1984cosmological, 2021MNRAS.507.2885F} for all radio point sources, and use this to scale the fluxes to \ac{MWA} frequencies. This ensures consistent methodology between the two Stingrays and reduces uncertainties. 

We use this spectral index to scale the 944\,MHz \ac{EMU} point source flux densities to \ac{MWA} frequencies (`Scaled' column in Table~\ref{tab:scaled fluxes}). We then subtract this scaled flux density from the total measured flux density (\ac{MWA} rows in Table~\ref{tab:measured fluxes}) to estimate the diffuse flux density for the circle and tail regions at \ac{MWA} frequencies (`Diffuse' columns in Table~\ref{tab:scaled fluxes}). Errors are estimated as $\sim$10\% following our method for measured fluxes in Section~\ref{subsubsec:flux}.

We plot the estimated flux density from the diffuse emission against the observing frequencies to obtain a spectral index value for each object (Figure~\ref{fig:specindx}). We use the diffuse flux density measurements as described in Sec~\ref{subsubsec:flux} (the `Diffuse' columns in Table~\ref{tab:measured fluxes}) for the circular and tail regions. For the \ac{MWA} data the diffuse flux density used (`Diffuse' column in Table~\ref{tab:scaled fluxes}) is estimated as the total measured flux density (\ac{MWA} rows in Table~\ref{tab:measured fluxes}) minus the scaled point source contribution (`Scaled' columns in Table~\ref{tab:scaled fluxes}. The \ac{WALLABY} data is only used for the circle region of Stingray~1 as the tail region could not be reliably measured. These diffuse flux densities are used to fit the spectral indices for the circle and tail components ($\alpha_{\text{Circle}}$ and $\alpha_{\text{Tail}}$ in Figure~\ref{fig:specindx}). The circle and tail diffuse flux densities are then summed together to calculate the total diffuse flux density for the entire area of each Stingray at each frequency (excluding the \ac{WALLABY} 1368\,MHz data point as the tail flux density could not be measured). {\bf When we defined the regions in Section~\ref{subsubsec:morphology}, we ensured that the circle and tail regions were adjacent regions but not overlapping, thus summing these flux densities together does not measure any emission twice.} These total flux density values for the diffuse emission are then used to calculate the spectral index of the whole object ($\alpha_{\text{Total}}$ in Figure~\ref{fig:specindx}).

The spectral indices were calculated from the slope of the line of best fit through the data points, which is calculated using the \textsc{linregress}\footnote{\url{https://docs.scipy.org/doc/scipy/reference/generated/scipy.stats.linregress.html}} function in the Python \textsc{scipy} package \citep{2020SciPy-NMeth}. This method applies a linear least-squares regression method to the data to find the line of best fit, and the quoted uncertainty is the standard error of the fit \citep{theil1950rank}. The plotted spectral indices are shown in Figure~\ref{fig:specindx} and the calculated final values are shown in Table~\ref{tab:scaled fluxes}.

We note that the quoted spectral index uncertainties are the statistical uncertainties, and this method of estimating extragalactic spectral indices could introduce additional uncertainties into this method which are not reflected in the standard error. This uncertainty can be seen in the fit of some of the data points, such as the tail section of Stingray~1 (Figure~\ref{fig:specindx}, left), which shows deviation from the linear fit. This uncertainty is most pronounced in the Stingray~1 tail, most likely due to this section having more point source contribution than the other areas. Thus, the flux density scaling assumption has more impact here, and this may cause deviation from a linear fit.

As this method relies on assuming a spectral index of $\alpha=-0.7$ for the radio point sources, it is possible that this value overestimates the steepness of the derived spectral indices. 
While some extragalactic radio sources can have significantly steeper spectral indices (e.g. ultra-steep spectrum sources), this would result in an overestimation of the point source contribution at \ac{MWA} frequencies and the spectral indices may be flatter.

This method assumes a spectral index which is scaled to \ac{MWA} frequencies and removed, introducing some uncertainty to the measurements. This is particularly relevant for the source WISEA J012939.26–535841.0 located at the tip of tail of Stingray~1. At \ac{MWA} frequencies, this source begins to dominate and we are unable to separate the diffuse tail emission. This source is also catalogued as the \ac{GLEAM} source J012938$-$535829 with a measured spectral index of $\alpha\,=\,-1.24\pm0.07$~\citep{2017MNRAS.464.1146H} using the \ac{GLEAM} bands. Due to the larger beam size, these \ac{MWA} measurements are likely contaminated by the diffuse tail, and thus cannot be used to separate the emission. These uncertainties make it difficult to determine exact values for the spectral indices, however the different methods applied all give relatively steep spectral indices, indicating that the Stingrays are overall non-thermal emitters with $\alpha\lesssim-0.8$. Additionally, the circular area of Stingray~1 and the tail area of Stingray~2 have been catalogued as GLEAM sources GLEAM J012911$-$535238 with $\alpha\,=\,-1.32\pm0.13$ and GLEAM J024626$-$564007 with $\alpha\,=\,-1.86\pm0.10$~\citep{2017MNRAS.464.1146H}. These values are solely from the \ac{MWA} bands, but provide further evidence for the Stingrays's non-thermal nature.

\paragraph{Spectral Index Map: }

We also generate a spectral index map using the \ac{ASKAP} \ac{EMU} and \ac{WALLABY} data for Stingray~1 (Figure~\ref{fig:specindexmap}). We are only able to generate a spectral index map for Stingray~1, as there is not yet \ac{WALLABY} data available for Stingray~2 and so we do not have sufficiently high-resolution radio data at multiple frequencies to resolve the Stingray~2 spectral index structure adequately. The total spectral index map including point sources is shown on the left, and only the diffuse component is shown on the right. To map the diffuse component, the point sources were removed using \ac{AeReS} as described in Section~\ref{subsubsec:flux} and the two largest point sources were manually masked. Both \ac{ASKAP} images were convolved to a common resolution of 30\arcsec\ before the spectral index map generation. The map was generated using the \textsc{maths} function from the \ac{MIRIAD} software package~\citep{Sault1995}. As we are calculating the spectral index value from only two datasets, we are able to do it with a simple equation where we calculate $\alpha$ as the linear slope between two data points: $\alpha$\,=\,($\log (S_{\text{WALLABY}})-\log (S_{\text{EMU}})$)/($\log (1367)-\log (944))$, where $S$ are the pixel values measured from the \ac{WALLABY} and \ac{EMU} images respectively. The \textsc{maths} function generates this spectral index value for each pixel in the \ac{ASKAP} images to generate the spectral index map. The cuts were selected at the highest contour level that showed the full extent of Stingray~1 in the convolved images so as to reduce background contamination. Figure~\ref{fig:specindexmap} (left) has the 944\,MHz \ac{ASKAP} radio contours overlaid to show the Stingray's extent. The point sources present in this image predominantly show values $\alpha\,\sim$\,--0.7, providing justification for our earlier assumption.

Figure~\ref{fig:specindexmap} (right) shows a steep spectral index for Stingray~1, with a circle average of $\sim$--2.4 and a tail average of $\sim$--2.8. 
This is steeper than the estimated values from Figure~\ref{fig:specindx} where the \ac{MWA} scaled fluxes were also accounted for. This is not unexpected as Stingray~1 is difficult to accurately measure in the \ac{WALLABY} images due to the lower sensitivity, shorter observation time, smaller bandwidth, and the intrinsically lower flux density at the higher frequency. There is also possibly a significant error in both estimation methods, due to the scaling of the \ac{MWA} point source flux densities, and due to calculating the spectral index from only two frequency measurements. Despite these uncertainties however, it is clear that Stingray~1 appears as a steep spectrum source, likely indicating an evolved object, as discussed in Section~\ref{sec:discussion}.

\begin{figure*}
    \centering
    \includegraphics[width=1\linewidth]{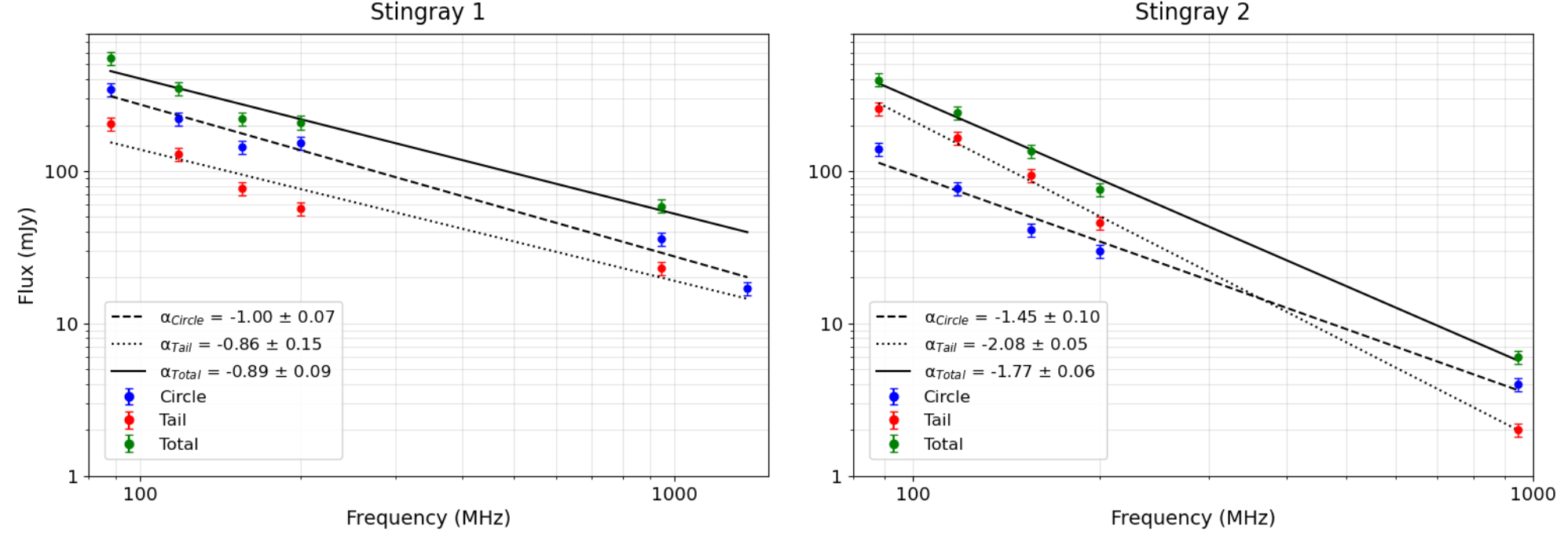}
    \caption{Spectral index components for each Stingray. The different colours and line styles represent the different regions. Flux density values used are the diffuse values from Tables~\ref{tab:measured fluxes} and~\ref{tab:scaled fluxes}. The line of best fit is calculated using the linear least-squares regression method.}
    \label{fig:specindx}
\end{figure*}

\begin{figure*}
    \centering
    \includegraphics[width=\textwidth]{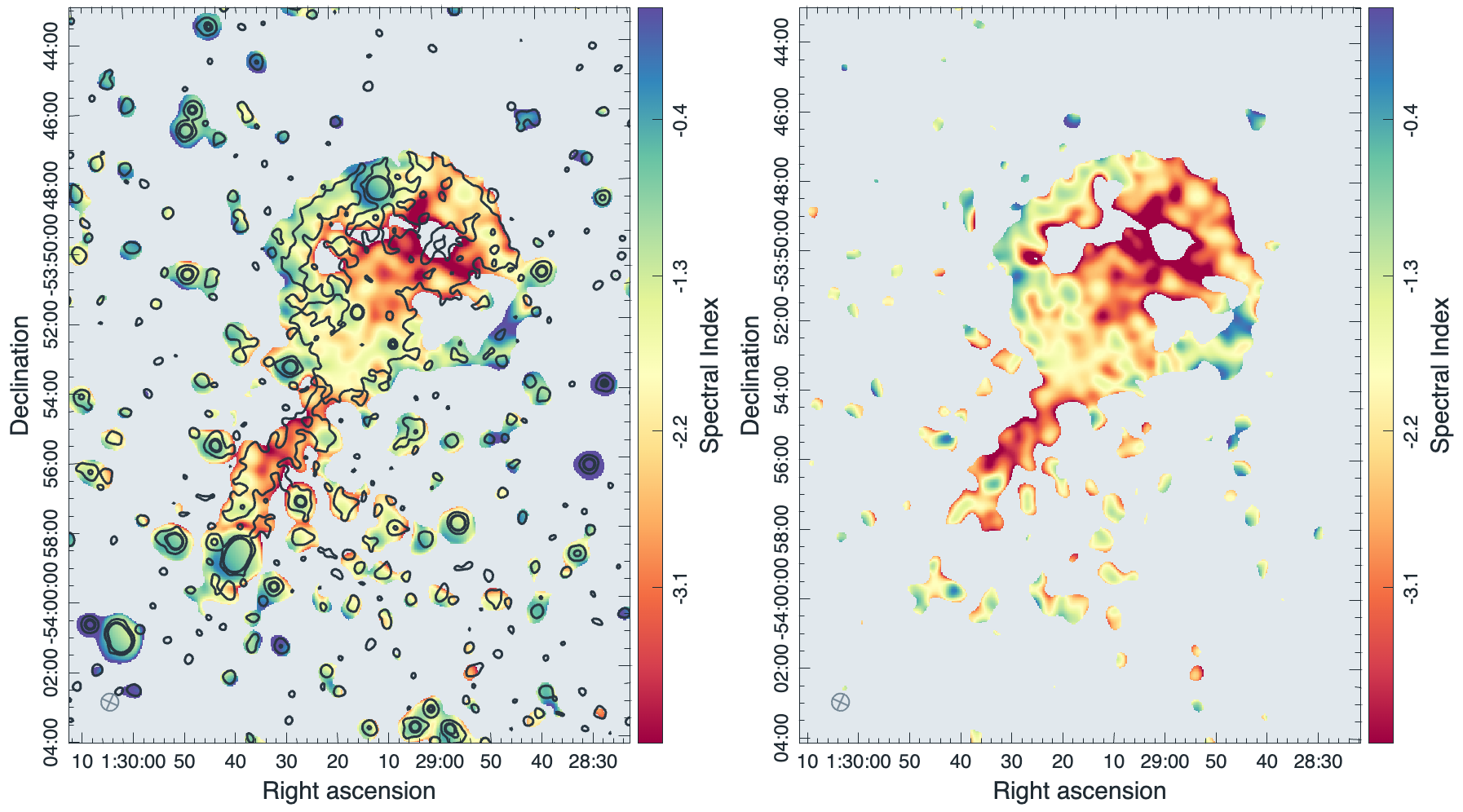}
    \caption{
    Spectral index maps of Stingray~1 derived from \ac{ASKAP} images at 944 and 1368\,MHz after each was smoothed to a resolution of 30\arcsec. The resolution is shown in the bottom left corner of each image. {\bf Left: } Spectral index with point sources included. Black contours are from 944\,MHz \ac{ASKAP} images at levels of 60, 300, and 400\,$\mu$Jy beam$^{-1}$. {\bf Right: } Spectral index map of the diffuse emission component after point source subtraction using AeReS. The two largest point sources were manually masked before image generation.}
    \label{fig:specindexmap}
\end{figure*}

\subsubsection{Surface brightness}
\label{subsubsec:surface brightness}

The surface brightness is a measure of how bright the radio emission is per unit of angular area of the source. The relationship is given as $\Sigma_{1\text{GHz}}=S_{1\text{GHz}}/\Omega$, where $\Sigma_{1\text{GHz}}$ = surface brightness at 1\,GHz, $S_{1\text{GHz}}$\,=\,flux density at 1\,GHz, and $\Omega$\,=\,angular area of object \citep{book1}. We use the measured spectral indices from Table~\ref{tab:scaled fluxes} to calculate the scaled flux at $\nu$ = 1~GHz and calculate the area of the source from the regions described in Section~\ref{subsubsec:morphology}. We use these values to calculate the radio surface brightness of the Stingrays (Table~\ref{tab:surface brightness}).

These are particularly low surface brightness values and the causes for this differ for each origin scenario discussed. More detail is given for each possible origin scenario in Section~\ref {sec:discussion}. This lower value could also indicate expansion into an extremely rarefied environment, and this scenario would also help explain the observed symmetry in the circular region. 
For extragalactic sources, such a low surface brightness may be caused by redshift dimming. This is an effect where sources at higher redshifts experience dimming in their surface brightness due to their large distance \citep{Calvi_2014}.

\begin{table*}
    \centering
    \caption{Calculated radio surface brightness for each Stingray. Area is calculated from regions defined in Section~\ref{subsubsec:morphology} and flux density is scaled to $\nu$\,=\,1\,GHz using spectral indices from Table~\ref{tab:scaled fluxes}.}
    \vskip.25cm
    \begin{tabular}{|c|ccc|} \hline
     & $\Omega$ & $S_{\text{1\,GHz}}$ & $\Sigma_{\text{1\,GHz}}$ \\
     & ($\times$10$^{-6}$\,sr) & (mJy) & ($\times$10$^{-23}$\,W m$^{-2}$ Hz$^{-1}$) \\ \hline
     Stingray~1 & 4.0 & 524 & 13.1 \\
     Stingray~2 & 1.9 & 5.2 & 2.7 \\ \hline
    \end{tabular}
    \label{tab:surface brightness}
\end{table*}

\subsection{\HI\ analysis}
\label{subsec:HI}

A commonly used method to determine distances to celestial objects is that of \HI\ absorption. There are regions of \HI\ gas and dust throughout the Universe, and these clouds absorb and emit light at specific wavelengths as the light travels through it. This absorption occurs at a specific redshift depending on the object's velocity, which can be attributed to a specific distance from us. These absorption dips can be seen in an object's \HI\ spectrum, and by measuring the velocity of the absorption dips, the distance to the objects can be constrained. A good theoretical review is given in~\citet[]{leahy2010distances} and this \HI\ absorption method has been used extensively for both Galactic and extragalactic sources \cite[e.g.,][]{koribalski1995hi, Leahy2012, 10.1093/mnras/stab754}.

\begin{figure*}
    \centering
    \includegraphics[width=0.9\linewidth]{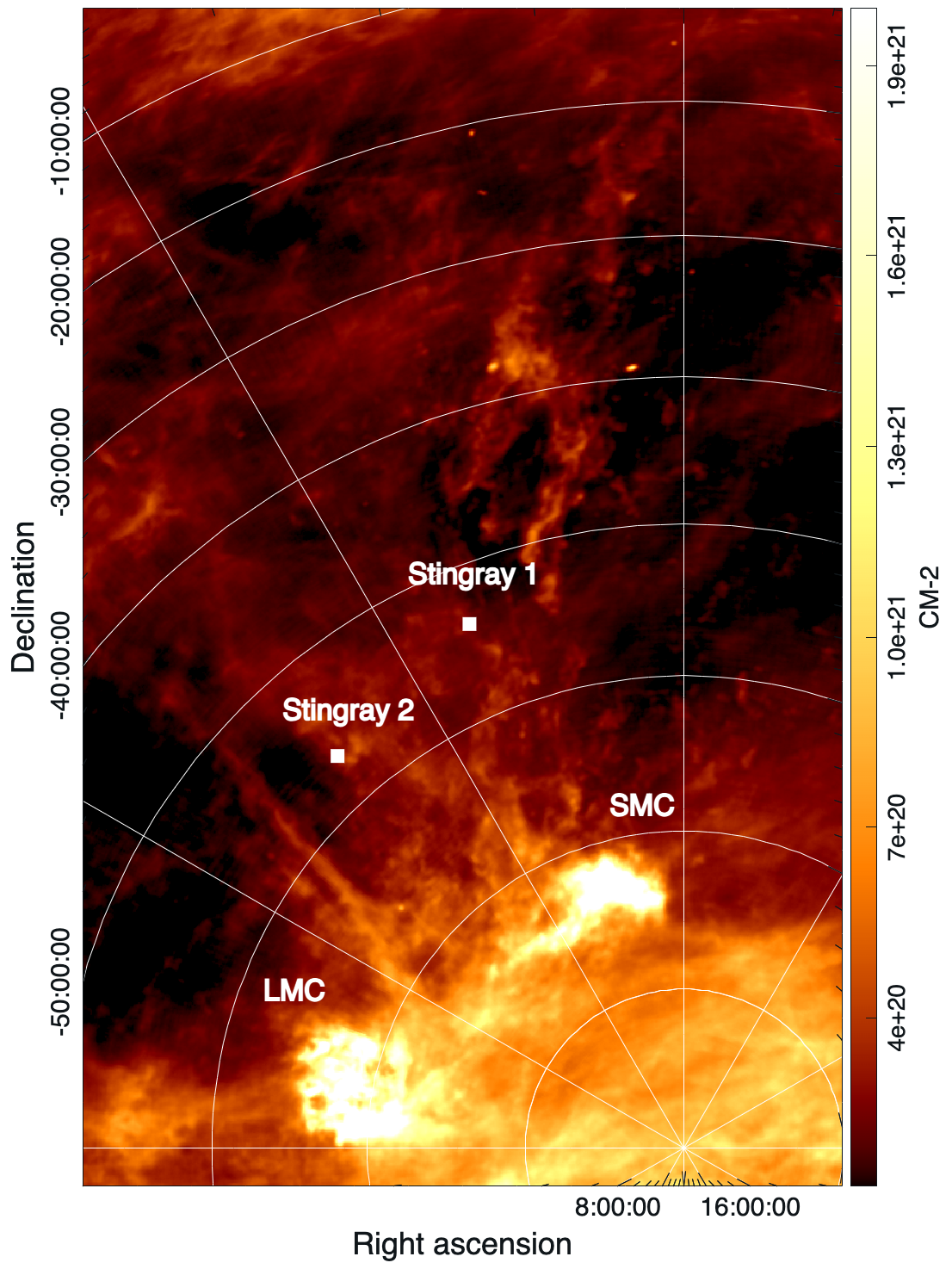}
    \caption{Total \HI\ column density map of the Magellanic System using data from the Parkes Galactic All-Sky Survey \citep[GASS,][]{HI4PI}. The locations of the LMC and SMC as well as Stingrays~1 and 2 are annotated. The Magellanic Stream extends north (upward) of the SMC and consists of several \HI\, filaments and diffuse emission.}
    \label{fig:HImap}
\end{figure*}

By measuring absorption in an object's \HI\ spectrum, we can determine if an object is located in front of or behind these \HI\ clouds, allowing us to constrain the distance if the cloud distance is known.
Several surveys have been conducted to determine the geometry, velocity, and \HI\ gas distribution within the \ac{MS} \citep{1979AJ.....84.1173H, Putman_2000, blandhawthorn2001galactic, 2003ApJ...586..170P, Bruns2005}. The \ac{MS} is typically split up into 4 sections with differing velocities. We compare the locations of the Stingrays with the \HI\ map of~\citet[their Figure~2]{Putman_2000} and determine both Stingrays are located in Section~MS~I (Figure~\ref{fig:HImap}). This region represents the positive velocity section that are above Galactic velocities, that is \mbox{30\,km\,s$^{-1}< V_{\text{LSR}} < 250$\,km\,s$^{-1}$}. We should also see absorption from Galactic \HI, and these velocities correspond with \mbox{--40\,km\,s$^{-1}< V_{\text{LSR}} < 30$\,km\,s$^{-1}$ \citep{Bruns2005}}.

We measure the \HI\ absorption spectrum for the Stingrays (Figure~\ref{fig:HI}) using data from the HI4PI survey (Section~\ref{HI4PI}) using the same total region as defined in Section~\ref{subsubsec:morphology} and Table~\ref{tab:summary}. Both show absorption dips at the expected velocities of the Milky Way and \ac{MS} (Figure~\ref{fig:HI} and Table~\ref{tab:HIresults}).

\begin{figure*}
    \centering
    \includegraphics[width=1\linewidth]{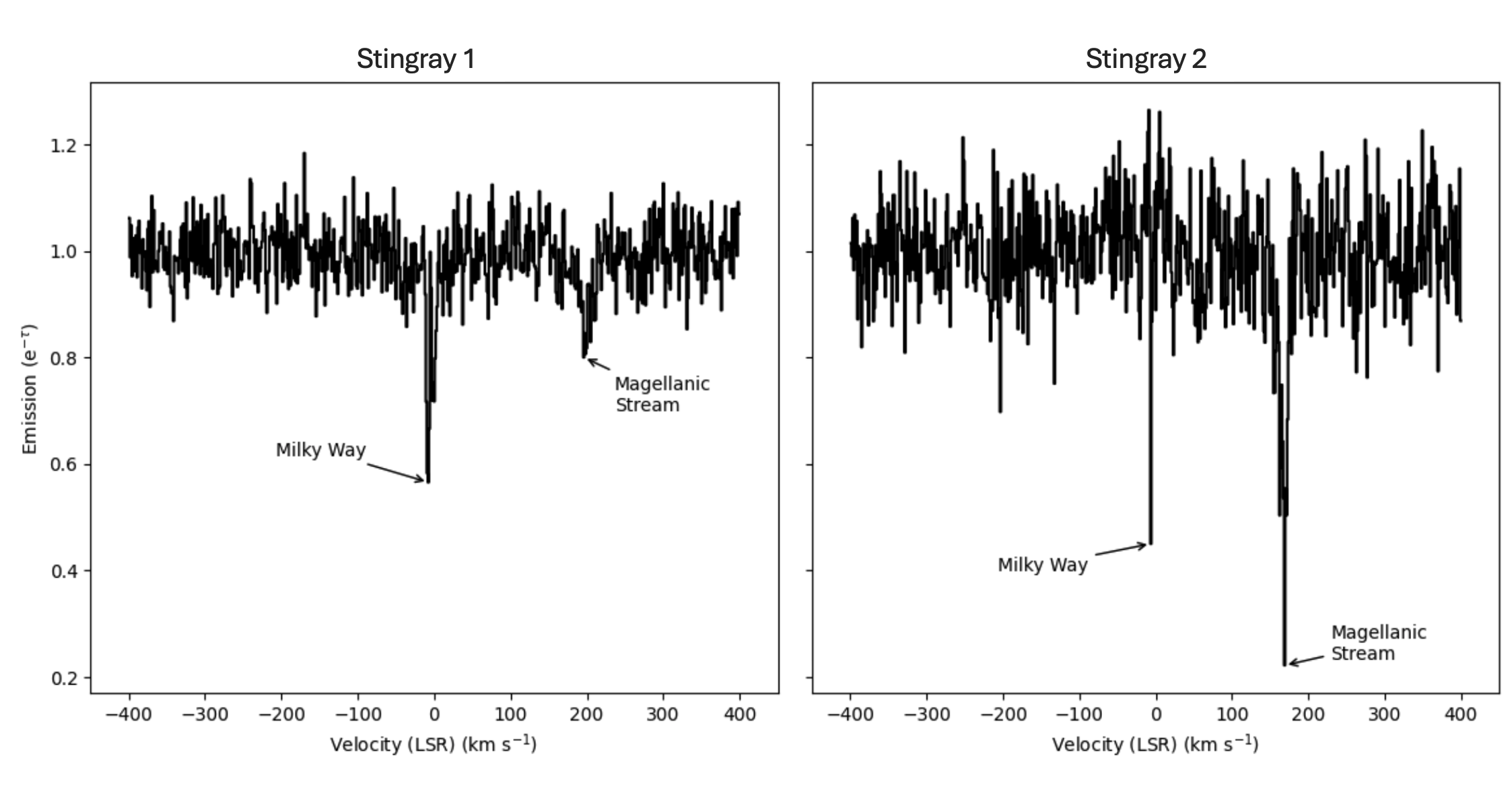}
    \caption{\HI\ absorption spectrum for each Stingray
    from HI4PI data. The absorption dips for both the Milky Way and Magellanic Stream are annotated.}
    \label{fig:HI}
\end{figure*}

We select an area that is stable with no obvious dips in both spectra as an estimate of the background noise levels. Taking this region as \mbox{$-400\leq V_{\text{LSR}}\leq-100$\,km\,s$^{-1}$} we measure the standard deviations and the signal-to-noise ratio of our peaks (Table~\ref{tab:HIresults}). We measure background noise levels of 0.05 for Stingray~1 and 0.08 for Stingray~2. The background noise level for Stingray~2 is $\sim$1.5 times that of Stingray~1. This could be caused by difficulties isolating the diffuse emission from the background due to the significantly lower surface brightness (Table~\ref{tab:surface brightness}). Despite the higher background noise, we still obtain a spectrum with clear absorption dips.

\begin{table*}
    \centering
    \caption{Results from \HI\ absorption analysis. $\sigma$ is calculated from the background region, taken as $-400\leq V_{\text{LSR}} \leq-100$\,km\,s$^{-1}$.}
    \vskip.25cm
    \begin{tabular}{|c|c|ccc|ccc|} \hline
         & $\sigma$ & \multicolumn{3}{c|}{Milky Way peak} & \multicolumn{3}{c|}{Magellanic Stream peak} \\ \cline{3-8}
         &  & $V_{\text{LSR}}$ (km s$^{-1}$) & $e^{-\tau}$ & signal to noise & $V_{\text{LSR}}$ (km s$^{-1}$) & $e^{-\tau}$ & signal to noise \\ \hline
         Stingray~1 & 0.05 & --7.6 & 0.6 & 8.3 & 195.9 & 0.8 & 3.8 \\
         Stingray~2 & 0.08 & --6.3 & 0.5 & 7.0 & 168.9 & 0.2 & 9.8 \\ \hline
    \end{tabular}
    \label{tab:HIresults}
\end{table*}

Both Stingrays show observable dips at \ac{MW} and \ac{MS} velocities. Stingray~1 shows a \ac{MW} peak at $V_{\text{LSR}}$\,=\,$-$7.6\,km s$^{-1}$ with a value e$^{-\tau}$\,=\,0.6 at a 8.3$\sigma$ detection level, and a \ac{MS} peak at $V_{\text{LSR}}$\,=\,195.9\,km s$^{-1}$ with a value e$^{-\tau}$\,=\,0.8 at a 3.8$\sigma$ detection level. Stingray~2 shows a \ac{MW} peak at $V_{\text{LSR}}$\,=\,$-$6.3\,km s$^{-1}$ with a value e$^{-\tau}$\,=\,0.5 at a 7.0$\sigma$ detection level and a \ac{MS} peak at $V_{\text{LSR}}$\,=\,168.9\,km s$^{-1}$ with a value e$^{-\tau}$\,=\,0.2 at a 9.8$\sigma$ detection level.

We take any point above a 5$\sigma$ detection to be a real detection and any point above 3$\sigma$ to be a probable detection. As expected, both Stingrays show real absorption dips for the Milky Way. We also detect a real absorption dip for Stingray~2 for the \ac{MS}, and a probable dip for Stingray~1 corresponding with the \ac{MS} velocities. This analysis shows that both Stingrays are located beyond the \ac{MW} as we clearly detect absorption due to the \ac{MW} \HI. Stingray~2 also shows clear absorption in \HI\ at the \ac{MS} velocities, indicating that Stingray~2 is located behind the \ac{MS}. Stingray~1 shows a probable absorption at \ac{MS} velocities, indicating that it is likely located behind the \ac{MS}, but the detection is not conclusive.

\section{Discussion}
\label{sec:discussion}

We investigate two main origin scenarios of the Stingrays, the 
first being that of a Galactic, or near Galactic, source, where the origin scenarios discussed are a supernova remnant (SNR) from a runaway star(s) within the \ac{MS} (see Section~\ref{subsec:runaway SNRs}), a circumgalactic \ac{SNR} on the outskirts of the \ac{MW} or \ac{MS} (see Section~\ref{circumgalactic SNRs}), and a Galactic pulsar-wind nebula (PWN) (see Section~\ref{subsubsec:PWN}. The second origin scenario is that of a distant extragalactic source where the origin scenarios discussed are radio active galactic nuclei (AGN) (see Section~\ref{subsubsec:AGN}),  dying radio galaxies (see Section~\ref{dying radio galaxy}), galaxy clusters (see Section~\ref{subsubsec:galaxy cluster}), galaxy pairs/ groups (see Section~\ref{subsubsec:galaxy group}), head-tail radio galaxies (see Section~\ref{subsec:headtail radio galaxy}), and Odd Radio Circles (ORCs) (see Section~\ref{subsubsec:ORC}). 

\subsection{Origin: Galactic/Near Galactic}
\label{galactic/near galactic}

It is possible that the Stingrays are associated with the \acf{MS} or that they are Galactic objects. We consider both scenarios here; specifically runaway \acp{SNR} from the \acfp{MC}, circumgalactic \acp{SNR} located on the outskirts of the \acp{MC} or \acf{MW}, and Galactic \acp{PWN}. We also discuss the likelihood of the Stingrays being associated with the \ac{MS}.

\subsubsection{Runaway SNRs}
\label{subsec:runaway SNRs}

One possible explanation for the unusual morphology of the Stingrays is that they are \acp{SNR} from runaway stars. Runaway stars are stars that have been ejected from their parent cluster, typically with high velocities. There are two main mechanisms that cause this stellar ejection; the ejection by gravitational interactions in dense star clusters, or when a star in a binary system undergoes a \ac{SN} explosion ejecting its companion \citep{1961BAN....15..265B,Filipovic2022}. Runaway stars have been detected from both the \ac{LMC} and \ac{SMC}, and several are confirmed to be massive enough to explode as Type~II~\ac{SN} \citep{Gvaramadze2011,2023ApJ...952...64L}. Hydrodynamic models have predicted that these runaway \ac{SN} may form asymmetric and unusual \ac{SNR} morphologies \citep{2006ApJ...649..779V, 10.1093/mnras/stv898,Filipovic2022}. This scenario could explain the unusual observed morphology and identification of \acp{SNR} within the \ac{MS} would help to further classify the \ac{MS} stellar population.

\subsubsection{Association with Magellanic Stream}
\label{subsubsec:association with MS}

Our \HI\ analysis shows a real Milky Way absorption dip for both Stingrays (signal to noise ratios: 8.3 for Stingray~1 and 7.0 for Stingray~2), a real \ac{MS} dip for Stingray~2 (signal to noise ratio: 9.8) and a probable \ac{MS} dip for Stingray~1 (signal to noise: 3.8). This indicates that we are seeing absorption from the \ac{MW} \HI\ for both Stingrays's emission indicating they are located outside of the Milky Way, and that we are seeing absorption from the \ac{MS} \HI\ for Stingray~2, indicating that Stingray~2 is also located beyond the \ac{MS}.
The spectrum for Stingray~1 is less definitive. Since we see a dip of $>3\sigma$ it is likely a real detection, and the lower absorption level may be due to the light travelling through a smaller amount of \HI. The \ac{MS} is not consistent in its \HI\ column density \citep{Bruns2005}, and it may be that this direction contains less \HI. This could be due to a lower \HI\ density in this region, or the \ac{MS} may be thinner in this direction so the emission is travelling through less material. It is also possible that Stingray~1 is located within the \ac{MS} instead of on the far side, and so the emission is travelling through less \HI. While it is difficult to discern between these possibilities without higher resolution \HI\ data, we can say that it is unlikely that Stingray~1 is located in front of the \ac{MS} as we are seeing a small amount of absorption at the \ac{MS} velocity distance. We therefore conclude that Stingray~1 has a possible but unlikely association with the \ac{MS}, and Stingray~2 has no physical association with the \ac{MS}.


\subsubsection{SNR spectral index}
\label{subsubsec:statistical comparison}

We compare the spectral indices of the Stingrays with those of the known \ac{LMC} and \ac{SMC} \ac{SNR} population using the statistical analyses of \citet{2017ApJS..230....2B,2024A&A...692A.237Z} for the \ac{LMC} and \citet{2019A&A...631A.127M,2024MNRAS.529.2443C} for the \ac{SMC}. Both analyses show similar distributions. 
The mean spectral index value is $\alpha\approx-0.5$, with a distribution range $-0.9<\alpha<0$. 

We now compare our measured spectral indices (Tables~\ref{tab:scaled fluxes} and~\ref{tab:summary}) with the theoretical \ac{SNR} average and the observed \ac{MC} distribution range. 
Both Stingrays have steeper spectral indices than the theoretical value, $\alpha=-0.5$~\citep{Bell1978}, as well as being outside of the \ac{MC} \ac{SNR} population. For Stingray~1, the spectral index, $\alpha=-0.89$, is on the extreme end of the distribution, and Stingray~2's spectral index, $\alpha=-1.77$, is far outside of the range entirely. This comparison, coupled with our \HI\ analysis, makes it unlikely that they are \acp{SNR} associated with the \ac{MS}.

It is possible that the Stingrays have flatter spectral indices than those estimated here, due to the inherent uncertainty in our scaling of the \ac{MWA} flux densities. If the extragalactic point sources instead have a steeper spectral index than we assumed, this would mean that we underestimated the point source contribution at \ac{MWA} frequencies, and thus the spectral indices may indeed fall within the \ac{MC} \ac{SNR} population range, although our \HI\ analysis (Section~\ref{subsec:HI}) still indicates that the objects are located beyond the \ac{MS}.


\subsubsection{Circumgalactic SNRs}
\label{circumgalactic SNRs}

There is the possibility that the Stingrays may be circumgalactic \acp{SNR}. These are \acp{SNR} that are located on the outskirts or just outside of their host galaxy. Two such \acp{SNR} have been previously observed near the \ac{LMC}, \ac{SNR} J0624--6948~\citep{2022MNRAS.512..265F,2025Sasaki} and \ac{SNR} J0614--7251~\citep{2025Sasaki}, so it is possible that the Stingrays are members of the \ac{MC} \ac{SNR} population that have moved outside of the Galaxy. 

The spectral index of the Stingrays ($\alpha\,=\,-0.89$ for Stingray~1 and $\alpha\,=\,-1.77$ for Stingray~2) is above the theoretical average of $\alpha\,=\,-0.5$ for \acp{SNR}. Stingray~2 is well outside of the \ac{MC} \ac{SNR} observed distribution ($-0.9<\alpha<0$) and Stingray~1 is just above the lower limit, making it unlikely that they are circumgalactic to the \ac{MS}. While the spectral index steepness may have been overestimated due to our assumptions about the \ac{MWA} point source flux densities, it is still likely that they fall outside of the \ac{MC} population, particularly in the case of Stingray~2, thus arguing against this classification. 

The size of the Stingrays may present issues with this possibility as well. If we assume an average distance of $\sim$55\,kpc to the \ac{MS} (50\,kpc for the \ac{LMC} and 60\,kpc for the \ac{SMC}), then the Stingrays have approximate physical diameters of 110\,pc (Stingray~1) and 80\,pc (Stingray~2), considering only the circular region. These diameters are larger than expected for the typical \ac{LMC} \acp{SNR} population, which~\citet{2023MNRAS.518.2574B} show to have mean diameters of 44.9\,pc (S.D. = 24.9\,pc), although there are a handful of known and candidate \acp{SNR} to have diameters $>$100\,pc \citep{2021MNRAS.500.2336Y, Veliki2025}. This \ac{LMC} distribution is similar to the statistical distributions of \ac{SMC} \acp{SNR}~\citep[mean diameter\,=\,48\,pc (S.D. = 19\,pc); ][]{2024MNRAS.529.2443C}, Galactic \acp{SNR}~\citep[mean diameter\,=\,21.9$\pm$1.7\,pc; ][]{2023ApJS..265...53R}, and \acp{SNR} in other nearby galaxies ~\citep[for example, M31 with a mean diameter\,=\,44.2$\pm$1.5\,pc; ][]{2023ApJS..265...53R, 2014SerAJ.189...15G}. While this larger size does not exempt the Stingrays from classification as an \ac{SNR}, the larger than expected size is evidence against this scenario, especially since this analysis only considers the size of the circle region, excluding the tail region.

\subsubsection{Parentless Pulsar Wind Nebula}
\label{subsubsec:PWN}

Finally, we could also consider a \ac{PWN} without an associated \ac{SNR} scenario, such as Potoroo~\citep{2024PASA...41...32L}, Lighthouse~\citep{2016A&A...591A..91P, 2014A&A...562A.122P}, or the Guitar nebula~\citep{1993Natur.362..133C}. While the Stingrays appear morphologically similar, the spectral index for such a \ac{PWN} is expected to be flat. This is in disagreement with what we find for the Stingrays, and so this scenario is deemed unlikely.


\subsection{Origin: Extragalactic}
\label{subsec:extragalactic}

Since our \HI\ analysis shows it unlikely that the Stingrays are located in the \ac{MS} or Galaxy, it is most likely that they are located on the same line of sight but in the background, far out from the Local Group. This is the second main scenario we investigate; that of distant extragalactic radio sources. 
The spectral index indicates non-thermal emission, and so we examine some classes of extragalactic objects that exhibit extended non-thermal radio emission; that includes radio \acp{AGN}, dying radio galaxies, galaxy clusters, galaxy pairs/groups, head-tail radio galaxies, and \acp{ORC}. A prudent next step in this analysis is to identify any known galaxies that may be a host galaxy.

There are multiple galaxies that appear in the NED database within the region of the objects. 
For Stingray~1, we find 16 galaxies in the region of diffuse emission, and for Stingray~2 we find 11 galaxies. These galaxies are best seen in the optical, and we show their spatial correlation using optical data in the g, r, i, and z bands from the \ac{DESI} Legacy Survey DR10\footnote{\url{https://www.legacysurvey.org/}} \citep{2019AJ....157..168D} (Figure~\ref{fig:DESI}). We also search the available optical and \ac{IR} images by eye to identify the brightest of these galaxies for analysis in the following relevant sections.

\begin{figure*}
    \centering
    \includegraphics[width=0.7\linewidth]{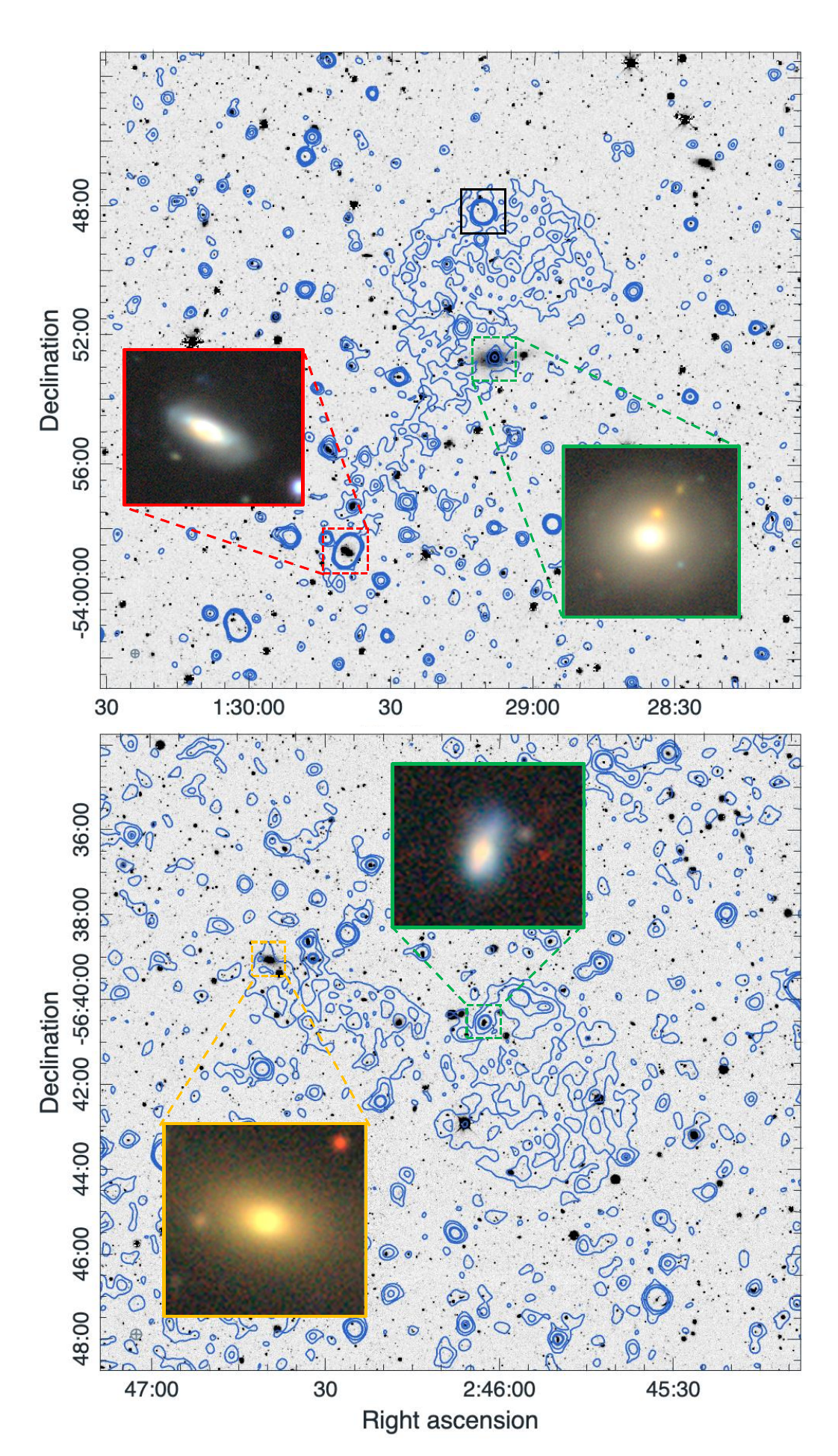}
    \caption{Optical images from the \ac{DESI} Legacy Survey DR10 overlaid with \ac{ASKAP} \ac{EMU} radio contours of Stingray~1 (top) and Stingray~2 (bottom). The optical images are averaged over the g, r, i, and z bands. The radio contours are at levels 60, 120, 200, and 400\,$\mu$Jy\,beam$^{-1}$ for Stingray~1  and at levels 30, 80, 200, and 250\,$\mu$Jy\,beam$^{-1}$ for Stingray 2 (right). The red inset shows potential \ac{AGN} galaxy host WISEA~J024639.50--563904.2 for Stingray~1 (top) analysed in Section~\ref{subsubsec:AGN}. The green insets show the coloured \ac{DESI}~DR10 images of the potential central galaxy groups/pairs, LEDA 425198 (top) and of 2dFGRS TGS845Z440 for Stingray~2 (bottom) analysed in Section~\ref{subsubsec:galaxy group}. The black square (top) shows the location of the galaxy cluster J012910.8$-$534812 for Stingray~1 and the orange square (bottom) shows the location of the galaxy cluster J024639.5$-$563904, with the inset showing the coloured \ac{DESI}~DR10 image of the BCG LEDA~398369 analysed in Section~\ref{subsubsec:galaxy cluster}.}    \label{fig:DESI}
\end{figure*}

\subsubsection{Radio AGN}
\label{subsubsec:AGN}

Extended non-thermal radio emission is observed in a fraction of \acp{AGN} ($\sim$10--20\%), known as radio loud \acp{AGN}~\citep{2020NewAR..8801539H, 10.1093/mnras/stac364}. These \acp{AGN} have highly energetic jets and lobes visible at radio frequencies due to charged particles emitted from the host galaxy's \ac{SMBH}. These jets can extend out to Mpc distances from the galaxy and form complex shapes~\citep{Blandford2019,2020A&A...642A.153D}. Radio \acp{AGN} exhibit a diverse variety of complex shapes (e.g.~\cite{pedlar1990radio, 1985ApJ...294L..85O, 1986ApJ...301..841O, 10.1093/mnras/stac2012, Velovic2023}). 

The jets and lobes of radio \ac{AGN} are caused by synchrotron emission from relativistic electrons and typically have steep spectral indices, $\alpha\sim-0.7$ to $\alpha\sim-1.0$, while the central core typically has a flat spectral index due to the constant energy input from the \ac{SMBH}~\citep{1984ARA&A..22..319B}. There is also a subset of \acp{AGN} with steeper spectral indices (up to $\alpha = -1.6$), known as dying or remnant radio galaxies~\citep{Murgia2011, Brienza2016}. The spectral indices of the Stingrays fall within these ranges, with Stingray~1 ($\alpha=-0.87$) being typical for \ac{AGN} jets, and Stingray~2 ($\alpha=-1.77$) (see Table~\ref{tab:summary}) being typical for a remnant radio galaxy. The spectral index alone is tentative evidence to use to classify the Stingrays, as we are currently unable to resolve any spectral differences between the possible jets/lobes and core.

The most typical morphology that appears similar to the Stingrays is that of an FR\,II~\citep{1974MNRAS.167P..31F} single-jetted \ac{AGN} with the host galaxy located at the end of the tail.
As radio \ac{AGN} require a host galaxy, we search for possible hosts in the optical and \ac{IR} regime, specifically DESI~DR10 and WISE catalogues, to search for the brightest galaxies within the Stingrays's areas. We primarily searched near the tail structure, as this emission appears most morphologically similar to typical radio \ac{AGN} jet appearance. We find a potential tail-located candidate for Stingray~1, WISEA J012939.26--535841.0 (Figure~\ref{fig:DESI}, top panel, red inset). This galaxy has a redshift of 0.059678$\pm$ 0.00015~\citep{1996ApJS..107..201L}, corresponding with a Hubble distance of 261.78$\pm$18.35\,Mpc. We are able to resolve the galaxy into three distinct components with the high-resolution \ac{EMU} image (Figure~\ref{fig:ASKAP}, top inset). The galaxy appears to be oriented perpendicular to the tail structure in the optical image (Figure~\ref{fig:DESI}), and the three radio components are oriented along the axis of the tail. 
If this is the host, then the diffuse emission is the jet and lobe structure of an \ac{AGN} jet originating from this galaxy.

We also note a potential candidate for Stingray~2, however, it is not located at the tail. The brightest emission section of the northern rim of Stingray~2 in the radio morphology appears to resemble a bent-tail radio galaxy. It is located on the periphery of the structure, although still within the Stingray~2 emission. We observe no optical counterpart in the \ac{DESI} optical images, and thus no identifiable host galaxy.



A possible concern with the radio \ac{AGN} scenario is that at these redshifts, the Stingrays would have vastly different physical sizes. Stingray~1 would measure $\sim$1.1\,Mpc from head to tip of the tail while Stingray~2 would measure $\sim$2.4\,Mpc. It is difficult to correlate such similar shapes with such a large difference in physical size unless Stingrays are some form of \ac{ORC} (see Section~\ref{subsubsec:ORC}). However, these large sizes may also indicate that they may be similar objects at different ages. If Stingray~2 is older, then this could help to explain the larger size, as well as the lower surface brightness and the steeper spectral index, as the emission has faded and lost its highest energy electrons as it ages.

While it is possible that the Stingrays are FR\,II single-jetted \acp{AGN}, the morphology still raises concerns with this classification. Mainly, there is a sharp disparity between the tail and circular sections, particularly pronounced for Stingray~1. This scenario would require a highly energetic radio jet to abruptly disperse and form an almost perfectly circular radio lobe. This would be an unusual morphology for an \ac{AGN}, as the lobes typically flare out into more asymmetrical shapes. However, \acp{AGN} can display a diverse range of morphologies, and so this scenario is a tentative possibility. 

\subsubsection{Dying Radio Galaxy}
\label{dying radio galaxy}

Another possibility is that the Stingrays are a relic or remnant emission of a dying radio galaxy. This is an evolutionary phase of some radio galaxies, where it is believed that the galactic nuclear activity has ceased, but the remnant radio lobes are still detectable for some time before they disappear completely \citep{10.1093/mnras/227.3.695, 1994A&A...285...27K}. Due to the cessation of constant energy injection, the lobes are subject to radiative losses of the relativistic electrons, which can result in steep radio spectral indices, $\alpha \lesssim $ --1.3 \citep{1994A&A...285...27K, 2023ApJ...944..176D}. The first relic radio galaxy discovered is IC\,2476 by~\cite{10.1093/mnras/227.3.695}, and while it is still a relatively sparse population, there are several other examples \citep{1993AJ....105..769H, Parma2007, 10.1093/mnras/stu360, Brienza2016, Duchesne_Johnston-Hollitt_2019, galaxies9040102}

The morphology of the Stingrays is also not particularly similar to any previously observed for dying radio galaxies, arguing against this scenario. It is also unusual that the morphologies would be so similar, while there is a substantial difference in spectral index. This difference in spectral index indicates that the objects may be of different ages, so it is unusual that radio galaxies at different ages display such similar morphologies. It is difficult to correlate these discrepancies with the scenario of a restarted remnant galaxy, but the morphology alone does not preclude the possibility.

A dying radio galaxy is a more plausible scenario for Stingray~2 compared with Stingray~1 due to the observed spectral indices. The spectral index of Stingray~2 ($\alpha_{\text{Total}}$\,=\,--1.77$\pm$0.06) (Tables~\ref{tab:scaled fluxes} and~\ref{tab:summary}), is within the range for a dying radio galaxy. The morphology of a connected circular and tail region is unusual for a dying radio galaxy, yet there are some that do exhibit unusual morphologies~\citep{1993AJ....105..769H, 10.1093/mnras/stu360}. If the Stingrays are relic radio galaxies, then it is likely that the host galaxy is one of those mentioned in Secs.~\ref{subsubsec:AGN} and~\ref{subsubsec:galaxy group}. In the case of the tail situated galaxy for Stingray~1, the emission would have formed by a jet forming a circular lobe-like structure, and in the case of the central situated galaxies, then it may be two jets with vastly different morphologies due to different environments or galactic winds. In the case of Stingray~1, we also have the possibility that the tail may represent a radio bridge connecting the two galaxies. This is a possibility as the tail-situated galaxy (described in Sec.~\ref{subsubsec:AGN}) and the central-situated galaxy (described in Sec.~\ref{subsubsec:galaxy group}) have similar redshift values and may be located in the same galaxy group. 

It has also been observed that dying galaxies can restart~\citep{2001AJ....122.2940J, 2009BASI...37...63S}, and so the dying radio lobe can be associated with spatially separated newly-born jets. Figure~\ref{fig:specindx} shows that the components of Stingray~2 have different spectral indices, indicating that the radio emission may be caused by different physical mechanisms or that they may be similar objects but with different ages. This is consistent with what may be expected from such a restarted source, where we only see one of the restarted jets. However, it is unexpected that the jet would display a steeper spectral index than the dying radio lobe. This is because the jet would have more efficient particle acceleration than the fading lobe, and so would exhibit a flatter spectral index. 

\subsubsection{Galaxy cluster}
\label{subsubsec:galaxy cluster}

Galaxy clusters can generate extended radio emission in the form of radio halos, typically circular emission regions around the cluster centre, and radio relics, elongated shapes at the cluster edges. Both are caused by synchrotron radiation from relativistic electrons, and are typically associated with clusters that display merger activity \citep{2019SSRv..215...16V, 2023A&A...672A..28L, Koribalski-Veronica2024}. Radio halos typically have spectral indices $\alpha=-1.2$ to $\alpha=-1.7$, while radio relics have steeper indices up to $\alpha=-2.0$ due to their older electron population \citep{2012A&ARv..20...54F, 2019SSRv..215...16V}. Galaxy clusters can have both radio halos and radio relics associated with them \citep[e.g.,][]{Pearce_2017,2023A&A...672A..28L, Velovic2023, 2024PASA...41...50M}.


We search several available galaxy cluster catalogues \citep{2016A&A...594A..27P, 2019ApJS..240...33G, 2021ApJS..253....3H, 2024A&A...685A.106B, 2024ApJ...967..123T, 2024ApJS..272...39W} and find one catalogued galaxy cluster for each Stingray, both identified in the catalogue of \citet{2024ApJS..272...39W}. For Stingray~1, this galaxy cluster is J012910.8$-$534812, located on the north-western edge of the Stingray~1 emission (see Figure~\ref{fig:DESI}). This galaxy cluster has a measured cluster redshift of $z_{\rm phot} = 0.9969$, a radius of 402\,kpc, and a mass of 0.56$\times$10$^{14}~M_\odot$. At this redshift, Stingray~1's total length would have a physical size of 6.72\,Mpc, 
much larger than the given cluster radius. This size is unrealistically large for such an association, and strongly argues against this scenario for Stingray~1. The cluster has a bright radio counterpart, which appears as a point source with an integrated flux density of 17.6\,mJy from the \ac{EMU} data. It likely corresponds to a known radio galaxy, WISEA J012910.84$-$534811.7 (also SUMSS J012910$-$534811)
and has a measured SUMSS 843\,MHz flux density of $S = 21.6\pm1.1$ mJy and a redshift of $z_\text{phot} = 0.94\pm0.06$.

For Stingray~2, the galaxy cluster is J024639.5$-$563904 
located at the tip of the tail section (see Figure~\ref{fig:DESI}). 
\citet{2024ApJS..272...39W} give a redshift of $z_{\rm phot} = 0.2127$, a radius of 773\,kpc, and a mass of 1.69$\times$10$^{14}~M_\odot$. At the cluster redshift, Stingray~2 would have a total physical length of 2.07\,Mpc, almost three times larger than the cluster radius from \citet{2024ApJS..272...39W}. 
The brightest cluster galaxy LEDA~398369 has a redshift of $z_\text{phot} = 0.222\pm0.005$ from the Legacy Survey DR9 dataset. 
LEDA~398369 may 
be responsible for the Stingray emission. \\

Galaxy clusters can have associated diffuse radio emission, e.g., halos and/or relics.
The clusters identified above are located within the Stingray emission, yet very near the edge in both cases. The diffuse radio emission from the Stingrays also extends much beyond the catalogued cluster radii. For Stingray~1, the emission is approximately 17 times larger than the cluster radius, and for Stingray~2, it is approximately 2$-$3 times larger. Galaxy cluster emission typically ranges from a few hundred kpc up to 1$-$2\,Mpc in size \citep{vanWeeren_2017}. Stingray~1 emission is significantly larger than this range, and Stingray~2 is on the very upper edge. This larger size, combined with the galaxy cluster not being located near the geometric centre of the emission, makes it unlikely that these are hosts for the emission. It is possible that there are non-catalogued galaxy clusters within the emission which may make these larger sizes more realistic. Due to the Stingray's large sizes however, the potentially uncatalogued cluster would have to be relatively nearby. For example, an upper redshift limit of $\sim$0.3 would give the Stingrays physical sizes of $\sim3.8\,$Mpc for Stingray~1 and $\sim2.7$\,Mpc for Stingray~2, which are unrealistically large for a galaxy cluster scenario. Therefore, the cluster would need to be $z<0.3$ as a conservative upper limit, and this would require several tens or hundreds of nearby galaxies to be missing from current catalogues, which is an unlikely scenario.

As both catalogued clusters are on the edge of the emission, if there were another cluster at the other end or more centrally located, then the emission may represent a bridge structure connecting these separate structures. This may help explain the unusual morphology and the larger sizes, however we currently have no direct evidence that there are more clusters located within the emission.

Another caveat to this scenario is that galaxy cluster radio emission is typically accompanied by X-ray emission due to the hot, ionised gas. The area of the Stingrays has only been observed by eROSITA in X-ray, not by the more sensitive {\it Chandra} or {\it XMM-Newton} telescopes. We searched in the available eRASS images, and detected no corresponding diffuse X-ray emission for the Stingrays, potentially arguing against this galaxy merger scenario. It is possible, there may be X-ray emission that is below eROSITA's sensitivity limits and would require deeper X-ray observations to be detected.

\subsubsection{Galaxy pair/group}
\label{subsubsec:galaxy group}

There is also the possibility that the radio emission may be being caused by a smaller galaxy group, or an interacting galaxy pair, which is the \ac{BGG} for a small galaxy group. 
We searched the entire emission by eye for optical sources, with a preference for more centrally located sources, which appeared to have a potential interacting companion.

For Stingray~1, a possible host is the early-type galaxy WISEA~J012908.17--535241.1 (LEDA~425198)}, shown in Figure~\ref{fig:DESI} (top panel, green inset) which has a spectroscopic redshift of $z_{\text{spec}} = 0.051873\pm0.000150$
\citep{2004MNRAS.355..747J, 2009MNRAS.399..683J}. 
This would give Stingray~1 a distance of 227.8$\pm$16.0~Mpc, a total size of$\sim$0.843~Mpc (the longest axis from the edge of the circular region to the tip of the tail), and a circular size of $\sim$0.425\,Mpc. Just west of LEDA~425198 is a smaller companion galaxy (LEDA~425213) 
which has a similar redshift of $z_\text{phot} = 0.066\pm0.008$ from the Legacy Survey DR9 \citep{2025MNRAS.536.2260Z}. The diffuse light surrounding the two galaxies suggest they are an interacting galaxy pair.

While galaxy pairs do not typically generate complex diffuse radio emission, it has recently been observed in the ``Physalis'' system \citep{Koribalski2024-Physalis}. Such complex emission is also observed in slightly larger galaxy groups such as ``Stephan's Quintet'' \citep{Stephan1877, Xu2003}.
To determine if there is a sufficient galaxy density in the area surrounding LEDA 425198 to support a larger group, we sample a 1\D\ region around Stingray~1 using the NED database. We find 69 galaxies with known redshift values within the region. We take a redshift range which contains LEDA 425198, 
0.031873 $<\,z\,<$ 0.051873, and find that 21 of these fall within. Therefore, this galaxy density makes it a possibility that there is a small galaxy group where LEDA 425198 is the \ac{BGG}.

LEDA 425198 is located in the centre of Stingray~1, between the circular and tail sections 
with a possible interacting companion, LEDA~425213 (see Fig.~\ref{fig:DESI}, top panel, green inset). It is possible that the distinct circle and tail sections we see are tidal features caused by this potential interaction. They may be outflows from the elliptical galaxy that show distinctly different shapes due to asymmetric gravitational effects from the interaction. That is, the circular region may be an outflow that is being bent into this circular shape by the gravitational interaction, causing this distinct morphology. A potential issue with this scenario is the large physical size at this redshift. While galaxy clusters can reach up to a few Mpc in some cases, galaxy group or pair emission is typically smaller. For example, the Physalis structure has a physical size of 145$\times$116\,kpc \citep{Koribalski2024-Physalis} and Stephan's Quintet has a slightly larger length of $\sim$0.6\,Mpc \citep{2022Natur.610..461X}. Such a large physical size is unlikely to be caused by a small galaxy group or interacting galaxy pair. Therefore, LEDA~425198 may be a \ac{BGG} for a group here, however it is difficult to confirm this group membership.

If LEDA~425198 is the host galaxy, then the Stingray~1 emission may be bent remnant lobes from potential interaction with its companion. In this scenario, the remnant or relic lobes originate from previous activity of the currently inactive \ac{BGG} LEDA~425198, and the resulting lobes are being bent by the surrounding environment, with one being pushed down to form the linear tail-like feature, and the other bending around in an arc to form the circle-like structure.

There is a similar galaxy within Stingray~2 (Figure~\ref{fig:DESI}, bottom panel, green inset), WISEA~J024602.67--564033.5 (2dFGRS TGS845Z440), that may also have an interacting companion. For the companion, we follow a similar procedure for the Stingray~1 object, searching the Gaia DR3~\citep{GaiaDR32023}, 2MASS~\citep{Skrutski2006}, and allWISE~\citep{Cutri2014} catalogues, however do not find the source listed. Therefore, this may represent a galaxy pair, but this cannot be confirmed with the current data. Following a similar argument, this may be a \ac{BGG} host for Stingray~2. 2dFGRS TGS845Z440 has a redshift of 0.154600$\pm$0.000297, as measured by the 2dF Galaxy Redshift Survey~\citep{2003astro.ph..6581C}. At this redshift, Stingray~2 would have a distance of 682.4$\pm$47.8~Mpc and a size of $\sim$1.903~Mpc (the longest axis from the edge of the circular region to the tip of the tail). Using a similar analysis as above, we find 305 galaxies within a 1$^{\circ}$ diameter. This is a higher galaxy density than for Stingray~1 overall, and the redshift range that 2dFGRS TGS845Z440 is located in, $0.1499<z<0.1599$, contains 31 galaxies. Therefore, it is possible that, similar to Stingray~1, there is a small galaxy group with 2dFGRS TGS845Z440 as the \ac{BGG}.

Since 2dFGRS TGS845Z440 is also located roughly between the circular and tail sections and appears to have an interacting companion, we can make a similar argument as above. It is possible that the tail and circular region are both jets, with one being bent by the gravitational interaction. Stingray~2's physical size is even larger in this scenario, and thus it is equally unlikely that such a large structure could be caused by a a small galaxy group or pair. 
It is possible that both Stingrays display an interacting galaxy pair at a central location and that the unique morphologies may be the result of an unusual gravitational interaction. The main issue with this scenario is that it is difficult to explain the large physical sizes, making this explanation unlikely.


\subsubsection{Head-tail radio galaxy}
\label{subsec:headtail radio galaxy}

Another possibility is that the Stingrays could be head-tail radio galaxies, which are formed by the passage of a strong wind or shock front. This passage suggests blowing back one of the jets and making it almost disappear, which can generate plumes or ring-like structures. The underlying idea is presented in~\citet{2019ApJ...876..154N}, and this interpretation was used to explain the unusual structure of the Corkscrew galaxy~\citep{2024MNRAS.533..608K}. In the case of the Stingrays, this would involve jets originating from the tail-situated potential host galaxy. Head-tail radio galaxies have several predicted morphological characteristics which appear to match both Stingrays, particularly Stingray~1. They predict a one-sided jet ending in a circular structure at the end of the jet, where the wind is blowing the jet back to form a circular plume-like structure. They predict that the head section is slightly separated from the tail section, which matches with the slight brightness dip observed in Stingray~1 where the tail meets the circular region. It also predicts that the host galaxy is slightly offset from the jet origin point, which is observed in Stingray~1, where the obvious radio galaxy WISEA J012939.26--535841.0 (Figure.~\ref{fig:DESI}, top panel, red inset), is offset slightly to the east from the tail. This scenario plausibly explains all of the unusual morphological characteristics of Stingray~1 and thus presents the most likely scenario out of those discussed. 

This scenario is also a possibility for Stingray~2, which displays a similar morphology. The caveat here, however, is that there is no obvious host galaxy for the Stingray~2 emission. There is a prominent radio source, LEDA~398369 (Figure~\ref{fig:DESI}, bottom panel, red inset), located at the tail tip for Stingray~2, 
but this is classified as a cluster and not a single radio galaxy. While head-tail structures such as this are associated with single radio galaxies, it is possible that a similar structure could form from a similar mechanism on a larger scale, explaining Stingray~2's morphology. Conversely, it is possible that there is a similar host galaxy as for Stingray~1, but it has not yet been identified.


\subsubsection{ORC}
\label{subsubsec:ORC}

\acp{ORC} are a recently discovered class of radio sources whose nature and origin is still under investigation. Three single \acp{ORC} are currently known, \acp{ORC} 1 and 4~\citep{galaxies9040083} and \ac{ORC} 5~\citep{10.1093/mnrasl/slab041}, each centred on a massive elliptical galaxy. They have sizes of $\sim$1\arcmin, or 300--500\,kpc at the host galaxy redshift. The \ac{ORC} 2+3 pair~\citep[Macgregor et al. in prep]{Norris2021ORC}, consist of a radio ring, without a central galaxy, and a diffuse blob. These are most likely the lobes of a re-started radio galaxy.

The known single \acp{ORC} are characterised by their edge-brightened, near circular radio emission, for which no counterparts have been detected at non-radio wavelengths, and their steep spectral indices. They are often associated with a central host galaxy, and diffuse radio emission. MeerKAT images of \ac{ORC} 1 also show internal ring structures~\citep{2022MNRAS.513.1300N}. Various formation mechanisms have been suggested~\citep{galaxies9040083, 10.1093/mnrasl/slab041, 2023ApJ...945...74D, Shabala2024}, but the rarity of \acp{ORC} means our knowledge of their properties is still very limited. The search for \acp{ORC} is ongoing, and several \ac{ORC} candidates are discussed in the literature \citep[][Filipovi\'c et al. in prep, Macgregor et al. in prep]{Gupta2022, 2023MNRAS.520.1439L, Koribalski-Veronica2024}, as well as possibly related, much closer radio shell systems \citep{Koribalski2024-Physalis}.

The two peculiar radio sources discussed in this paper, Stingrays~1 and 2, share some common characteristics with \acp{ORC}. Their main body is predominantly circular, somewhat edge-brightened, partially filled with diffuse emission, and have a steep spectral index. However, no central radio source or obvious host galaxy was detected. Furthermore, a distinct, one-sided tail structure is detected, similar in size and surface brightness to the main body. Also, both Stingrays have angular sizes much larger than the currently identified \acp{ORC}. Therefore, it is possible that the Stingrays may be a type of \ac{ORC} that displays jet-like structures. \ac{AGN} jets viewed from a specific orientation have been theorised as a potential explanation for \acp{ORC}~\citep{ galaxies9040083, 2021PASA...38....3N, 2024ApJ...974..269L, Shabala2024}. The origin of \acp{ORC} is still debated in the literature, and jets are not defined as one of the characteristic observable features. Therefore, it would be premature to make this classification without further investigation due to these discrepancies. 

It is also possible that the morphology may be more similar to an \ac{ORC} if viewed from a different orientation. For example, if the host galaxy were at the tip of the jet and the orientation was such that the jet was directed along our line of sight and then expanded out into the circular region, it is possible that the jet would not be visible from this orientation, and instead, the Stingray would resemble a structure where the host galaxy appears to be inside a circle of diffuse emission. In this orientation, the Stingray may appear much more morphologically similar to an \ac{ORC}. 

Another potential possibility is that \acp{ORC} have jets when they are younger which fade as they age. Therefore, the population that we have identified thus far no longer has observable jets. In the case of the Stingrays, if the jets were to fade then what would be left behind is a quite circular patch of emission that would not be appear to be associated with a galaxy; a situation which holds true for some \acp{ORC}, namely the \ac{ORC} 2+3 pair~\citep[Macgregor et al. in prep]{Norris2021ORC}. As the true nature and origin of \acp{ORC} is still under investigation, these are currently purely speculative scenarios.

\subsubsection{Chance Alignment}
\label{subsubsec: chance}

It is also possible that the circular and tail regions of both Stingrays are not physically associated at all, and their unusual morphology is caused by a chance alignment of two or more sources. It should also be noted that both Stingrays display a slight drop in intensity at the region where the circle and tail sections intersect. This might indicate that the emission is not physically associated. 

If this scenario is the case, then the objects would consist of a radio tail structure and a circular region. The tail structure would be typical of an unresolved \ac{AGN} (Section~\ref{subsubsec:AGN}), and the circular region could be a galaxy cluster halo (Section~\ref{subsubsec:galaxy cluster}) or an \ac{ORC} (Section~\ref{subsubsec:ORC}). Stingray~2 displays several optically bright galaxies in the circular region (Figure~\ref{fig:DESI}) that may generate enough diffuse radio emission to form this circular halo structure. This scenario is less likely for Stingray~1 as there are far fewer optically bright galaxies present in the circular region.

\section{Conclusions}
\label{sec:conclusion}

We have conducted a radio analysis of two unusually shaped diffuse radio sources that we named ``Stingrays''. These Stingrays are diffuse regions of radio emission, characterised by a circular section with a connected tail-like section extending out. There is no corresponding diffuse emission found at any other frequency. We investigate several possible origin scenarios, both Galactic/near Galactic and extragalactic.



We explored several Galactic/near Galactic scenarios: 

- Runaway \acp{SNR} from the \acp{MC}: This scenario involves runaway stars from the \acp{MC}, which then exploded and formed \acp{SNR} within the \ac{MS}. Our \HI\ analysis shows that the objects are likely not associated with the \ac{MS} and are in the same line of sight. The measured spectral index values are also not consistent with the \ac{MC} \ac{SNR} population. This scenario is deemed unlikely.

- Circumgalactic \acp{SNR}: Similar to the runaway \ac{SNR} scenario, this scenario involves \acp{SNR} which formed on the outskirts of the \acp{MC} and are located just outside the galaxies themselves. The spectral index values are not consistent with the \ac{MC} \ac{SNR} population, and neither are the physical sizes at the \ac{MC} distances. This scenario is deemed unlikely.

- Parentless \ac{PWN}: A \ac{PWN} without an associated \ac{SNR}. The steep spectral index is at odds with this scenario, and this scenario is deemed unlikely.

We also explored several extragalactic scenarios for the Stingrays.

- Radio \ac{AGN}: This scenario involves powerful jets and lobes from a radio \ac{AGN} which have formed unusual morphologies, likely through some kind of environmental interaction. This is a more likely scenario for Stingray~1, as a most likely host is identified as WISEA J012939.26–535841.0, located at the tip of the tail. The spectral indices and sizes can be explained in this scenario, and it is deemed a possible scenario for both Stingrays.

- Dying radio galaxy: This scenario involves a radio galaxy where the galactic nuclear activity has ceased, but the remnant radio lobes are still visible. The morphology causes some issues with this scenario, but the spectral indices, particularly for Stingray~2, are consistent. This is deemed a possible scenario for both Stingrays.

- Galaxy cluster: We find a catalogued galaxy cluster located in the emission of each Stingray, and galaxy clusters can display extended radio emission. The clusters are located on the peripheries of both Stingrays and the associated physical sizes are larger than expected for a cluster scenario at the given redshifts. This scenario would be possible if more galaxy clusters are found within the emission in the future. This scenario is currently deemed unlikely for both objects.

- Galaxy pair/group: There are possible interacting galaxy pairs for both Stingrays, and the galaxy density in both areas is sufficient to support a small galaxy group. Galaxy pairs and groups can display extended radio emission, although the physical sizes of the Stingrays are larger than expected from such a system. This scenario is currently deemed unlikely for both objects.

- Head-tail radio galaxy: This scenario involves a head-tail radio galaxy, which are typically formed by strong \ac{AGN} jets which are pushed back by strong winds or shock fronts, and form round circular structures. Both Stingrays meet the morphological criteria for these objects, and Stingray~1 has a possible host identified. This scenario is deemed likely for Stingray~1 and possible for Stingray~2, with the caveat that the Stingray~2 host would have to be identified.

- ORC: This scenario involves \acp{ORC}, which are circular regions of diffuse radio emission, observed exclusively at radio frequencies. These objects typically have central elliptical host galaxies, which are not identified in the Stingrays. Their morphology is also not characteristic of an \ac{ORC}, primarily as they are not fully circular. This scenario is deemed unlikely for both objects.

- Chance alignment: This scenario suggests that the Stingrays may not be a single structure, but in fact a circular region superimposed with a tail region. Finding two objects with similar morphologies caused by such a superposition would be statistically unlikely, but is possible for either one or both of the objects. This scenario is deemed possible for both objects.


Overall, several of these scenarios could explain the Stingrays unusual shapes, however, further observations would be required to resolve these uncertainties. For example, more sensitive X-ray observations could detect any corresponding diffuse X-ray emission and help determine the physical properties of the emission, while more sensitive radio observations may reveal detectable polarisation to help map any magnetic field properties and variations.

The final conclusion is that Stingrays are diffuse extragalactic non-thermal radio sources, but their exact nature remains unclear. Their unusual morphology could be caused by orientation effects or complex environmental interactions in several of the proposed scenarios. The most likely scenario from the current data is that of head-tail radio galaxies. This scenario is most likely for Stingray~1, and possible for Stingray~2, but the host galaxy for Stingray~2 would need to be identified.



\begin{acknowledgement}

This scientific work uses data obtained from Inyarrimanha Ilgari Bundara / the Murchison Radio-astronomy Observatory. We acknowledge the Wajarri Yamaji People as the Traditional Owners and native title holders of the Observatory site. The \ac{CSIRO}’s \ac{ASKAP} radio telescope is part of the Australia Telescope National Facility\footnote{\url{https://ror.org/05qajvd42}}. Operation of \ac{ASKAP} is funded by the Australian Government with support from the National Collaborative Research Infrastructure Strategy. \ac{ASKAP} uses the resources of the Pawsey Supercomputing Research Centre. Establishment of \ac{ASKAP}, Inyarrimanha Ilgari Bundara, the \ac{CSIRO} Murchison Radio-astronomy Observatory and the Pawsey Supercomputing Research Centre are initiatives of the Australian Government, with support from the Government of Western Australia and the Science and Industry Endowment Fund. 

The DESI Legacy Imaging Surveys consist of three individual and complementary projects: the Dark Energy Camera Legacy Survey (DECaLS), the Beijing-Arizona Sky Survey (BASS), and the Mayall z-band Legacy Survey (MzLS). DECaLS, BASS and MzLS together include data obtained, respectively, at the Blanco telescope, Cerro Tololo Inter-American Observatory, NSF’s NOIRLab; the Bok telescope, Steward Observatory, University of Arizona; and the Mayall telescope, Kitt Peak National Observatory, NOIRLab. NOIRLab is operated by the Association of Universities for Research in Astronomy (AURA) under a cooperative agreement with the National Science Foundation. Pipeline processing and analyses of the data were supported by NOIRLab and the Lawrence Berkeley National Laboratory (LBNL). Legacy Surveys also uses data products from the Near-Earth Object Wide-field Infrared Survey Explorer (NEOWISE), a project of the Jet Propulsion Laboratory/California Institute of Technology, funded by the National Aeronautics and Space Administration. Legacy Surveys was supported by: the Director, Office of Science, Office of High Energy Physics of the U.S. Department of Energy; the National Energy Research Scientific Computing Center, a DOE Office of Science User Facility; the U.S. National Science Foundation, Division of Astronomical Sciences; the National Astronomical Observatories of China, the Chinese Academy of Sciences and the Chinese National Natural Science Foundation. LBNL is managed by the Regents of the University of California under contract to the U.S. Department of Energy. The complete acknowledgments can be found at https://www.legacysurvey.org/acknowledgment/.

We thank an anonymous referee for their detailed and constructive comments which greatly improved our paper.

\end{acknowledgement}


\paragraph{Data Availability Statement.}

All ASKAP data products are made publicly available in the CSIRO \ac{ASKAP} Science Data Archive (CASDA)\footnote{\url{https://research.csiro.au/casda}}. The MWA GLEAM data is publicly available from the MWA All Sky Virtual Observatory (ASVO)\footnote{\url{https://asvo.mwatelescope.org/}} and from SkyView Virtual Observatory\footnote{\url{https://skyview.gsfc.nasa.gov/current/cgi/titlepage.pl}}. The HI4PI data is publicly available from the CDS server\footnote{\url{https://cdsarc.u-strasbg.fr/viz-bin/qcat?J/A+A/594/A116}}.

\printendnotes

\bibliography{Stingrays}

\appendix

\section{WALLABY radio continuum image}

\begin{figure}
    \centering
    \includegraphics[width=1\columnwidth]{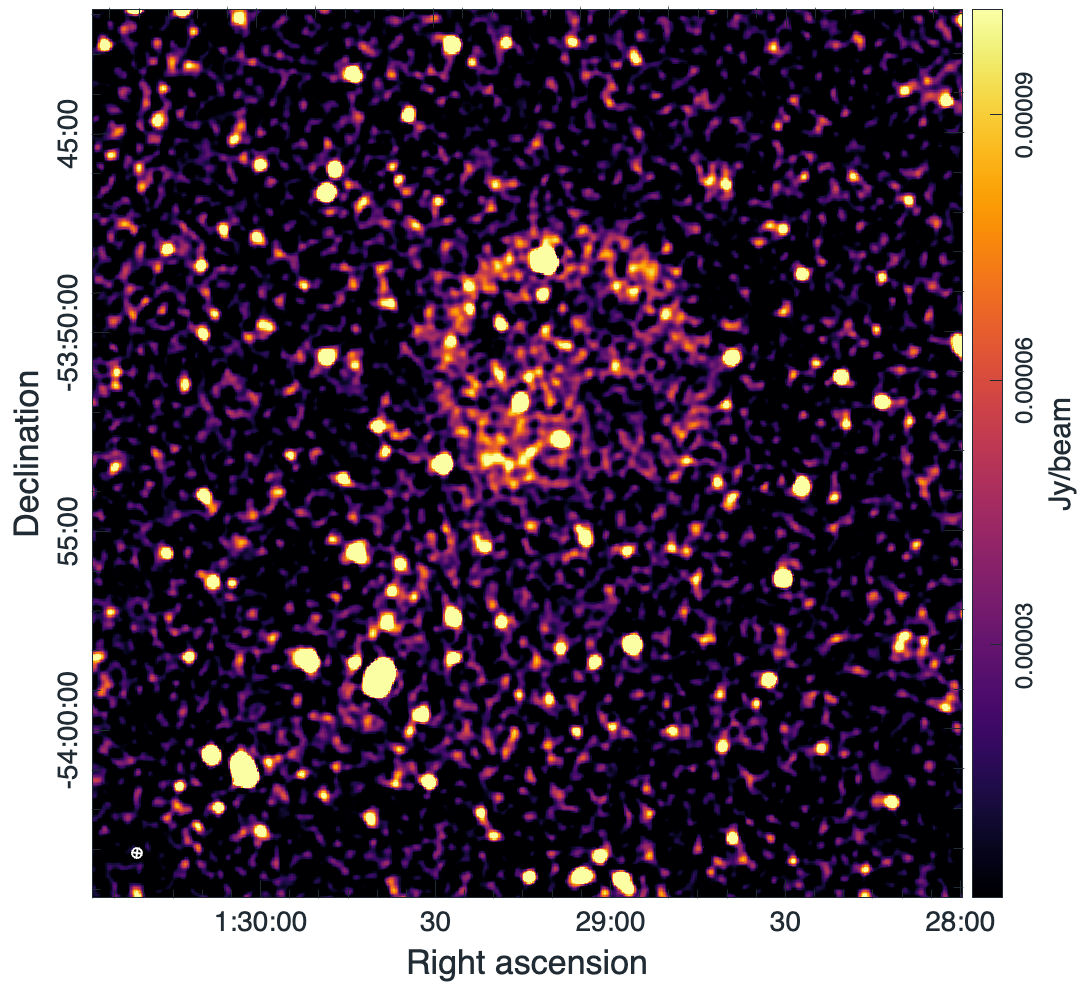}
    \caption{ASKAP WALLABY 1.4~GHz radio continuum image of Stingray~1. The image has linear scaling and has been convolved to 15\arcsec\ resolution, indicated in the bottom left corner. We measure an \ac{RMS} noise sensitivity of $\sigma \sim 25-30$~$\mu$Jy\,beam$^{-1}$ near Stingray~1.}
    \label{fig:A1WALLABY}
\end{figure}

\end{document}